\documentclass[12pt]{article}
\usepackage{listings}
\usepackage{semantic}
\usepackage{amsmath}
\usepackage{graphicx}
\usepackage{subfigure,adjustbox}
\usepackage[letterpaper, portrait, margin=1.2in]{geometry} 
\usepackage{flexisym}
\usepackage{bm}
\usepackage{float}
\usepackage{multirow}
\usepackage{booktabs}
\usepackage{tikz}
\usepackage{relsize}
\usepackage{caption}
\usepackage{color}
\usepackage{soul}
\usepackage{authblk}
\usepackage{blindtext}
\usepackage[font=small]{caption}
\usepackage{standalone}
\usepackage[round]{natbib}
\usepackage{lineno}
\usepackage{mathtools}
\usepackage{amsthm}
\usepackage{setspace}

\DeclarePairedDelimiter\floor{\lfloor}{\rfloor}

\setcitestyle{round,aysep={},yysep={;}}

\usetikzlibrary{shapes.gates.logic.US,trees,positioning,arrows}
\usetikzlibrary{trees}
\usetikzlibrary{arrows, decorations.markings}
\usetikzlibrary{intersections}
\usetikzlibrary{calc}

\tikzset{fontscale/.style = {font=\relsize{#1}}}
\tikzstyle{vecArrow} = [decoration={markings,mark=at position
   1 with {\arrow[]{open triangle 60}}},
   double distance=1.4pt, shorten >= 5.5pt,
   preaction = {decorate},
   postaction = {draw,line width=1.4pt, white,shorten >= 4.5pt}]
\tikzstyle{innerWhite} = [white,line width=1.4pt, shorten >= 4.5pt]

\begin{document}
\begin{flushright}
\end{flushright}
\begin{center}
{\Large Probabilities of unranked and ranked anomaly zones under birth-death models}\\

\bigskip


\bigskip

Anastasiia Kim$^1$, Noah A. Rosenberg$^2$, James H. Degnan$^{1*}$\\

$^1$Department of Mathematics and Statistics, University of New Mexico\\
$^2$Department of Biology, Stanford University\\
$^*$jamdeg@unm.edu\\
\end{center}

\clearpage

\section*{Abstract}
A labeled gene tree topology that is more probable than the labeled gene tree topology matching a species tree is called \textit{anomalous}. Species trees that can generate such anomalous gene trees are said to be in the \textit{anomaly zone}. Here, probabilities of \textit{unranked} and \textit{ranked} gene tree topologies under the multispecies coalescent are considered. A ranked tree depicts not only the topological relationship among gene lineages, as an unranked tree does, but also the sequence in which the lineages coalesce. In this article, we study how the parameters of a species tree simulated under a constant rate birth-death process can affect the probability that the species tree lies in the anomaly zone. We find that with more than five taxa, it is possible for species trees have both anomalous unranked (AGTs) and ranked (ARGTs) gene trees. The probability of being in either type of anomaly zones increases with more taxa. The probability of AGTs also increases with higher speciation rates.  We observe that the probabilities of unranked anomaly zones are higher and grow much faster than those of ranked anomaly zones as the speciation rate increases. Our simulation shows that the most probable ranked gene tree is likely to have the same unranked topology as the species tree. 
We design the software {\it PRANC} which computes probabilities of ranked gene tree topologies
given a species tree under the coalescent model.

\par
\section*{Introduction}
\par
\setlength{\parskip}{1em}
\par
In phylogenetic studies, gene trees are often used to reconstruct a species tree that describes evolutionary relationships between species. Gene trees that are contained within the branches of the species phylogeny represent the evolutionary histories of the sampled genes. 
The species tree is treated as a parameter, and gene trees are considered as random variables whose distributions depend on the species tree.

Probabilities of gene tree topologies in species trees have been studied for several decades \citep{nei1987,pamilo1988,Takahata1989,rosenberg2002,DegnanAndSalter05,Meng2009,Wu2012,Yu2012}, with an emphasis on unranked gene trees, gene trees in which the sequence of coalescences is not taken into account. For example, for the unranked gene tree $((A,B),(C,D))$,  the most recent ancestral gene of the $A$ and $B$ lineages could be either more or less recent than the most recent ancestral gene of the $C$ and $D$ lineages.  The probability of this unranked gene tree is calculated by summing both possibilities.
However, the probability distribution of the ranked gene tree topologies has also been derived, taking into account the temporal order of coalescence events \citep{Degnan2012,Stadler2012}.
In this case, we count as distinct the two gene trees $((A,B)_2,(C,D)_3)_1$ and $((A,B)_3,(C,D)_2)_1$, where the subscript indicates the ranking of the nodes. In the first of these two ranked gene trees, the $(C,D)$ coalescence, indicated by the largest subscript, is the most recent.

In 2006, Degnan and Rosenberg  defined the concept of an {\it anomaly zone}: a subset of branch-length space for the species tree in which the most likely unranked gene tree has a topology differing from the species tree topology. A non-matching gene tree topology that is more probable than the matching one was termed an {\it anomalous gene tree} (AGT) \citep{DegnanAndRosenberg06}.  An intuitive explanation for the existence of AGTs is that
when rankings of coalescences are not taken into account, gene trees that are more symmetric
can have more rankings than gene trees that are less symmetric \citep{DegnanAndRosenberg06, Rosenberg2013, xu2016challenges}. As an extreme case, a gene tree with only one two-taxon clade, called a {\it caterpillar}, can have only one possible ranking and can never be an AGT \citep{Rhodes}.

This explanation leads to a similar question for ranked trees: does the most probable ranked gene tree match the species tree? In the case of four taxa, this turns out to be the case: although caterpillar species trees can have unranked AGTs, they cannot have {\it anomalous ranked gene trees} (ARGTs), ranked gene trees that are more probable than the ranked gene tree with the same ranked topology as the species tree.  However, for five or more taxa, ARGTs do exist \citep{Degnan2012, degnan2012characterization, RosenbergandDisanto14}. The concept of anomalous gene trees has been further extended to consider anomalous unrooted gene trees (AUGTs) \citep{Unrooted2013}, in which unrooted gene trees that do not match the unrooted version of the species tree topology can be more probable than the matching unrooted gene tree. The concept of the anomaly zone can be even extended to phylogenetic networks \citep{InTheLight}.
In particular, a gene tree is anomalous if it is more probable than any gene tree displayed by the network. \citet{InTheLight} showed that three-taxon phylogenetic networks do not produce anomalies, but that symmetric phylogenetic networks with four leaves can produce anomalies.

Several properties of anomalous gene trees in different settings are known.
In particular, every species tree topology with five or more taxa produces AGTs \citep{DegnanAndRosenberg06, Rosenberg2013}.  The analogous result for unrooted gene trees is that every species tree topology with seven or more taxa produces AUGTs \citep{Unrooted2013}.
\citet{RosenbergandTao08} considered all sets of branch lengths that give rise to five-taxon AGTs. They found that the largest value possible for the smallest branch length in the species tree is greater in the five-taxon case (0.1934 coalescent time units) than in the previously studied case of four taxa (0.1568). This finding raises the question of whether species trees with more taxa are more likely to have AGTs. Studies for ARGTs \citep{degnan2012characterization} showed that neither caterpillar nor pseudocaterpillar species tree have anomalous ranked gene trees, where a {\it pseudocaterpillar} can be obtained from a caterpillar $(\dots(((A_1,A_2),A_3),A_4),\dots A_n)$ by replacing $(((A_1,A_2),A_3),A_4)$ with $((A_1,A_2),(A_3,A_4))$ \citep{rosenberg2007counting}. Strangely enough, although caterpillar gene trees cannot be AGTs, they can be ARGTs.  In addition, \citet{RosenbergandDisanto14} showed that as the number of species $n \rightarrow \infty$, almost all ranked species trees give rise to anomalous ranked gene trees.

\par
Evolutionary biologists have sometimes wondered how often anomalous gene trees arise in practice \citep{castillo2008factors,zhaxybayeva2009intertwined,Linkem2016}, since the existence of anomalous gene trees makes the method that chooses the most common gene tree as the estimate of the species tree statistically inconsistent in the anomaly zone.  A recent empirical identification of the anomaly zone is for gibbons \citep{shi2017coalescent}.
In spite of the many analytic results known about the various types of anomalous gene trees, less is known about how often they arise in practice. This question is difficult to answer because it requires some knowledge of the empirical distribution of branch lengths in the species trees.

To study the probability that the species tree lies in an anomaly zone, we examine random species trees generated from a constant rate birth-death process.  The approach we use is to simulate the species tree while computing gene tree probabilities analytically for each simulated species tree. This simulation can help to understand how often AGTs and ARGTs arise in practice, to the extent that birth-death processes are reasonable models for species trees and that we can understand typical birth-death process parameters. 
We additionally examine cross sections of anomaly zones to see how much overlap exists for different types of anomaly zones. This analysis shows that for larger trees, a species tree can simultaneously be in unranked and ranked anomaly zones.
 
We consider two types of gene trees: unranked and ranked gene trees. In general, we can compute the probability of an unranked tree topology from the probabilities of ranked gene tree topologies. The probability of an unranked gene tree topology can be obtained by summing the probabilities of all ranked gene tree topologies that share that unranked topology. We can therefore view unranked and ranked gene trees as preserving increasing amounts of information about the underlying rooted trees with full branch length information.

\par
This paper also introduces a computer program, \textit{PRANC}, for Probability of RANked gene tree topologies under the Coalescent model ({\tt https://github.com/anastasiiakim/PRANC}). The software computes probabilities of ranked gene trees given a species tree under the coalescent process. The program is implemented in \textit{C++} based on the approach proposed in earlier studies \citep{Degnan2012,Stadler2012}. 

We compute the probabilities of ranked and unranked gene tree topologies for all species trees with five to eight taxa to find a subset of speciation interval length space in which the species tree generates anomalous unranked and ranked gene trees. Studying the properties of anomalous gene trees, as well as examining connections between ranked and unranked anomaly zones, will help to find strategies for solving the problem posed during phylogenetic inference by the existence of anomalous gene trees.

\par
\subsection*{Definitions and notation}
\par
A species tree $\mathcal{T}$ is a binary tree with leaves that represent current species. We consider a rooted labeled ultrametric species tree with branch lengths given in coalescent units. For the rest of this paper, branch lengths in the species tree are in coalescent units unless otherwise stated. Here $1.0$ coalescent unit represents $N$ generations where $N$ is the effective number of gene copies. The same set of labels is used for both species and genes. In this article, all gene trees have one gene sampled per species.

\begin{figure}[]
\centering
\includegraphics[width=4.5in]{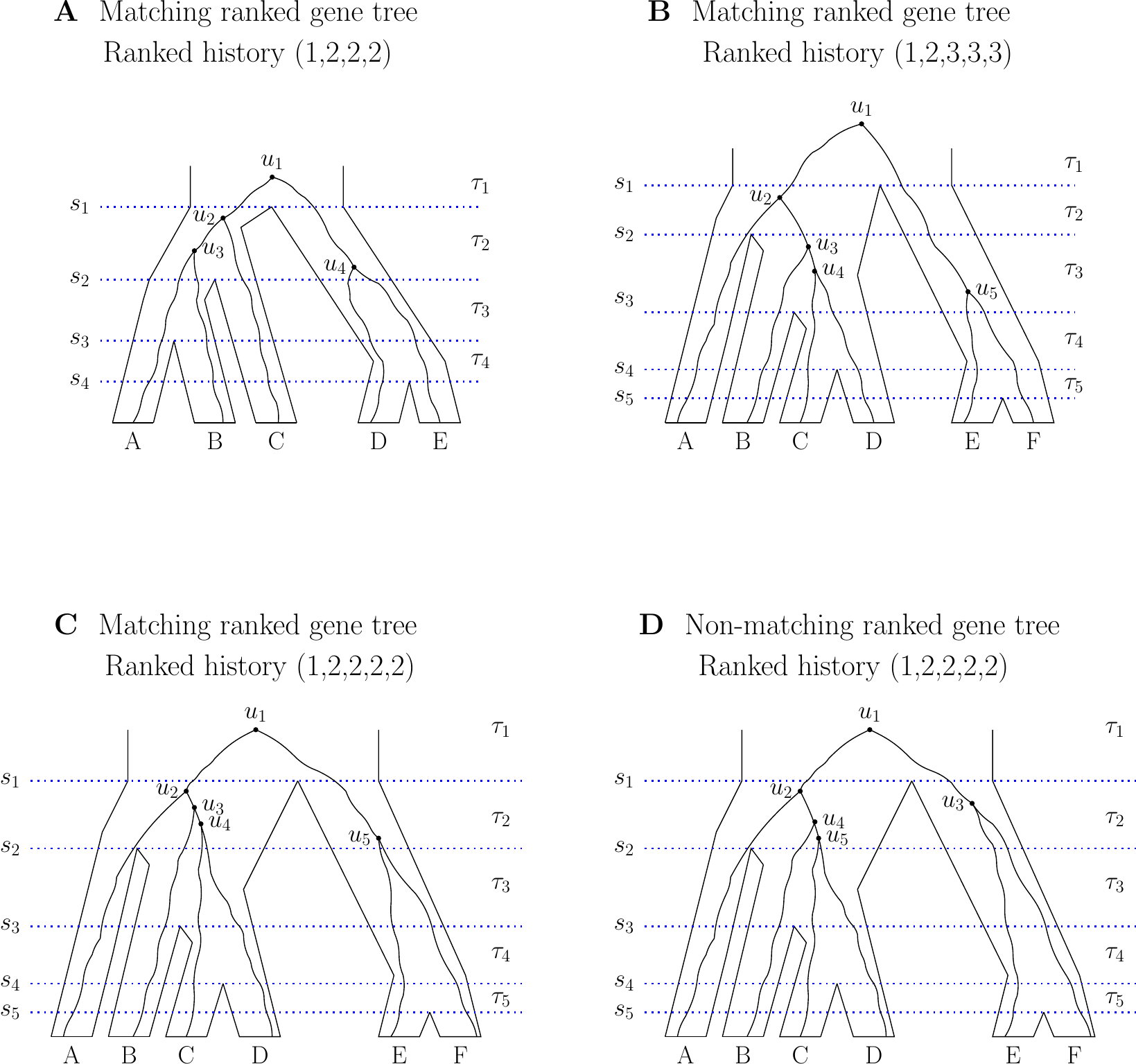}
\caption{Gene trees evolving on five-taxon (A) and six-taxon (B)--(D) species trees. The gene trees in (B)--(D) have the same unranked topology ((A,(B,(C,D))),(E,F)). Only the ranked gene tree topology in (D) does not match the ranked species tree topology. For each $i=1,2,...,n-1$, $s_i \geq 0$ denotes the time of the $i$th speciation, $\tau_i$ represents the interval between the $(i-1)$th and $i$th speciation events, and $u_i$ represents the $i$th coalescence (node with rank $i$) in the gene tree.  Interval $\tau_1$ has infinite length.}
\label{fig:genetrees}
\end{figure}

We assign ranks to the nodes of a species tree with $n$ labeled leaves according to their speciation order. Denote the time of the interior node of rank $i$ ($i$th speciation) by $s_i$, $i=1,2,...,n-1$. Time is zero for the leaves and increases going backwards in time: $s_1>s_2>...>s_{n-1}$, where $s_1$ is the time of the root (fig.~\ref{fig:genetrees}). For $i=2,3,...,n-1$, denote the interval between the $(i-1)$th and $i$th speciation events by $\tau_i$ and its length by $t_i =s_{i-1}-s_i$.

We write a ranked tree topology as a modified unranked tree topology using the Newick format, in which each clade is represented by a pair of parentheses, and we add a number after each clade to indicate its ranking. For example, the species tree in figure~\ref{fig:genetrees}A can be written $(((A,B)_3,C)_2,(D,E)_4)$. In the Newick format, we supress the labeling of the root node, which has rank $1$.

Let $\mathcal{G}$ be a ranked gene tree topology with the same labels for the leaves as species tree $\mathcal{T}$. Given a gene tree that evolves on a species tree $\mathcal{T}$, a $ranked$ $history$ can be defined as a non-decreasing sequence $x=(x_1,x_2,...,x_{n-1})$, where for $i=1,2,...,n-1$, $x_i=j$ if the $i$th coalescence occurs in species tree interval $\tau_j$ \citep{Degnan2012}. For example, in figure~\ref{fig:genetrees}B, the ranked history of the gene tree is $(1,2,3,3,3)$. One coalescence occurs in the species tree interval $\tau_1$, one in $\tau_2$, and three in $\tau_3$. We denote the probability under the coalescent model of a ranked gene tree topology with the particular ranked history $x$ by $P(\mathcal{G},x|\mathcal{T})$.

If a gene tree and species tree have the same unranked topology, then we describe the unranked topologies as identical and refer to the unranked gene tree as \textit{matching} the unranked species tree; otherwise, the gene tree topology is \textit{nonmatching}. Similarly, we say the ranked gene tree matches the ranked species tree if, and only if, they have the same ranked topology. At times we will also be interested in cases where a ranked gene tree has the same unranked topology as the species tree, meaning that if the ranks are ignored, the two trees are matching. Because the methods in this article involve only topologies of gene trees, the term ``gene tree" will be used to refer to the topology of the gene tree (without branch lengths) unless otherwise noted. Rooted labeled unranked or ranked gene tree topologies that are more probable than the labeled unranked or ranked gene tree topology matching the species tree are called anomalous gene trees and are termed AGTs and ARGTs respectively. Species trees that have unranked or ranked anomalous gene trees are said to be in the unranked or ranked \textit{anomaly zone} respectively.

\newpage
\section*{Results}
\par
\subsection*{Anomaly zones}
\par
We computed probabilities of ranked and unranked gene trees for species trees with five to eight taxa to find a subset of speciation interval length space in which a species tree has both anomalous unranked (AGTs) and ranked (ARGTs) gene trees. For plots comparing unrooted and unranked anomaly zones, see \citet{Unrooted2013}.

\par
\subsubsection*{Five taxa}
\par
Figure~\ref{fig:5taxa}A depicts a five-taxon species tree with interval lengths $t_2$, $t_3$, and $t_4$. The ranked topology shown is the only five-taxon species tree topology that possesses ARGTs. For fixed values of $t_4=0.05,0.075,0.1$, we computed the probabilities of all 105 unranked and all 180 ranked gene tree topologies on a grid with $t_2 \in [0.01, 3]$ and $t_3 \in [0.01,1]$. The anomaly zones were identified by finding the set of values of $t_2$, $t_3$, and $t_4$ for which at least one nonmatching unranked or ranked gene tree topology has probability exceeding the probability of the corresponding matching gene tree topology. 

Figure~\ref{fig:5taxa}B depicts slices of cross-sections of unranked and ranked anomaly zones for the five-taxon species tree in figure~\ref{fig:5taxa}A. For values of $t_2, t_3$ and $t_4$ considered, we observe that the unranked and ranked anomaly zones do not overlap for five-taxon species trees. As $t_4$ becomes smaller, the ranked anomaly zone increases in size, whereas the size of the unranked anomaly zone decreases. Although for the values of $t_i$ considered, we do not observe an overlap in unranked and ranked anomaly zones in the five-taxon case, these zones start to intersect for larger trees. 

\par
\subsubsection*{Six taxa}
\par
We next considered six-taxon trees. There exist six unlabeled tree shapes with six taxa. Excluding the caterpillar and pseudocaterpillar shapes, four of these, depicted in figure~\ref{fig:6taxontops2}, give rise to both AGTs and ARGTs. Figure~\ref{fig:6taxonzones} shows two-dimensional cross-sections of unranked and ranked anomaly zones for the six-taxon species tree topologies in figure~\ref{fig:6taxontops2}. For ease of visualization, we consider only two different values, denoted by $S$ and $L$, for the lengths of speciation intervals $t_i$. For each combination of $S \in [0.005,1]$ and $L \in [0.01,2]$, we computed the distributions of unranked and ranked gene tree topologies, and the presence of AGTs and ARGTs was then identified by comparing the analytical probabilities of the matching gene tree topology and the most probable nonmatching gene tree topology.

In the cases we examined, the two anomaly zones start to overlap only when lengths of the speciation intervals are short and not too distinct from each other. In particular, the intersection of anomaly zones is small for each topology, with the smallest overlap for the more balanced species tree topologies in figure~\ref{fig:6taxontops2}C and 3D.

\par
\subsubsection*{Seven and Eight taxa}
\par
We next sought to examine scenarios with seven and eight taxa (fig.~\ref{fig:8taxontops}) to determine if the interval-length cases giving rise to AGTs and ARGTs were similar to those seen in the case of six taxa.

The seven- and eight-taxon species trees were chosen so that they produce both AGTs and ARGTs. To find such topologies, we used a ``caterpillarization" technique of finding a short-short-long ($SSL$) pattern in three consecutive internal branches on a path from a tip to the root of the species tree, and setting all other branches to be long.  In \citet{Unrooted2013}, this technique was used to collapse taxa descended from long branches to be effectively a single taxon, making even a topologically balanced tree resemble a caterpillar when branch lengths are taken into account. More generally, the technique of setting some specific branches to be short and others to be long has been used frequently in identifying AGTs and ARGTs \citep{DegnanAndRosenberg06, Degnan2009properties, Degnan2012, degnan2012characterization, Rosenberg2013}.

\begin{figure}[H]
    \centering
    \mbox{}%
    \adjustbox{valign=T}{\subfigure{\textbf{A}}}
    \adjincludegraphics[valign=T,scale=1]{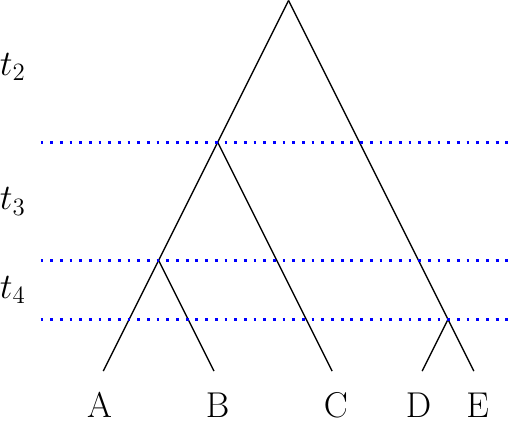}
    \label{subfigure:kinmorfirst}
    \hfill
    \adjustbox{valign=T}{\subfigure{\textbf{B}}}
    \adjincludegraphics[valign=T,scale=0.3, angle=270]{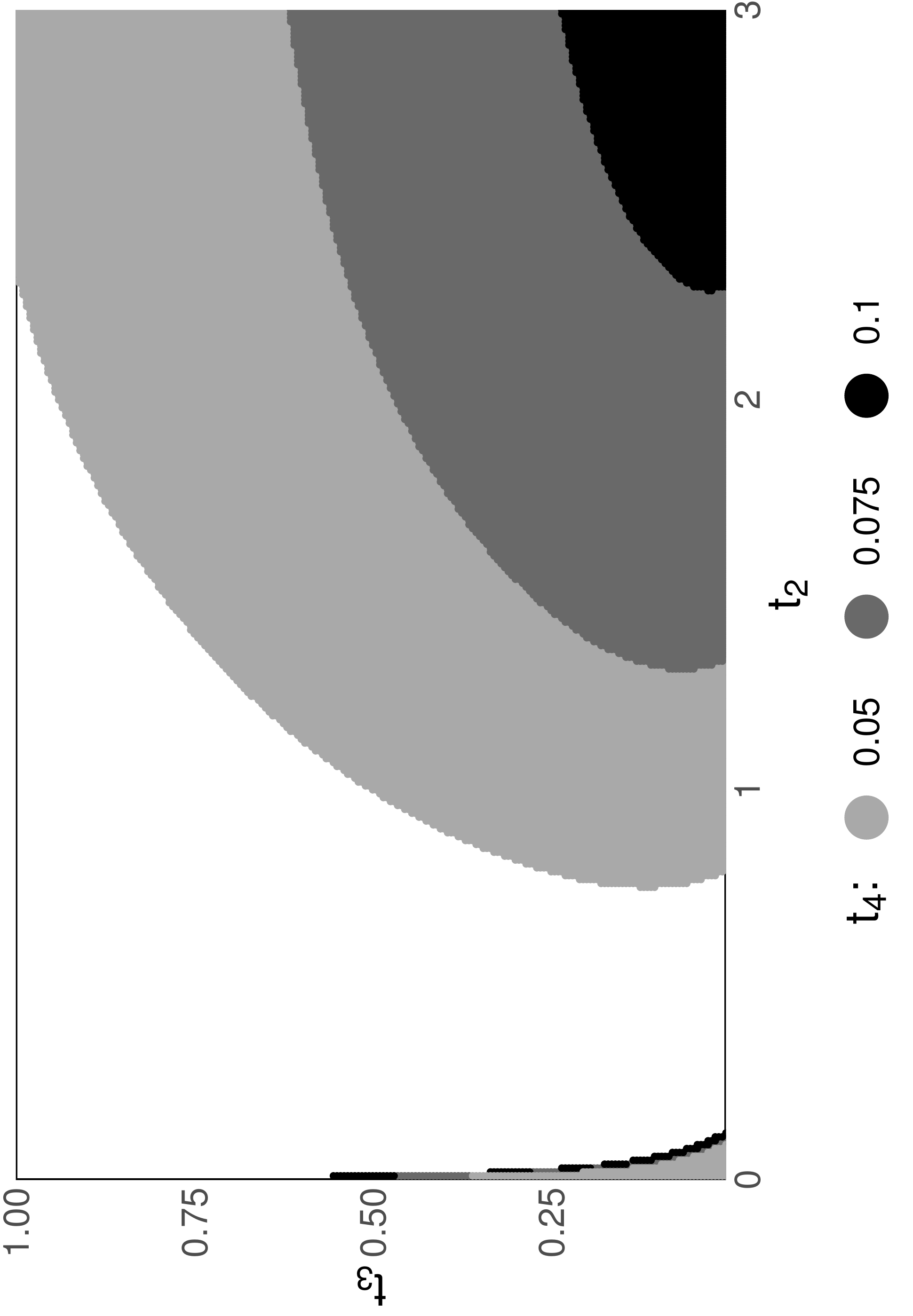}
    \label{subfigure:kindamgofirst}
\caption{Five-taxon anomaly zones. (A) The only ranked five-taxon species tree topology that produces ARGTs. The same species tree, with a gene tree evolving inside, is shown in figure~\ref{fig:genetrees}A. (B) Slices of the unranked (on the left side) and ranked (on the right side) anomaly zones for the topology in (A). For fixed values of $t_4$, each shaded region represents pairs of speciation interval lengths $(t_2,t_3)$ for which the most probable unranked (ranked) gene tree topology does not match the unranked (ranked) species tree topology. Each slice was generated by computing the probability distribution of gene tree topologies on a grid with $t_2 \in [0.01,3]$ and $t_3 \in [0.01,1]$, with increments of 0.01 for both variables. In the ranked case, the shaded region for a smaller $t_4$ contains the shaded region for a larger $t_4$. In the unranked case, the shaded region for a larger $t_4$ contains the shaded region for a smaller $t_4$.}  
\label{fig:5taxa}
\end{figure}

\begin{figure}[H]
\centering
\includegraphics[width=5in]{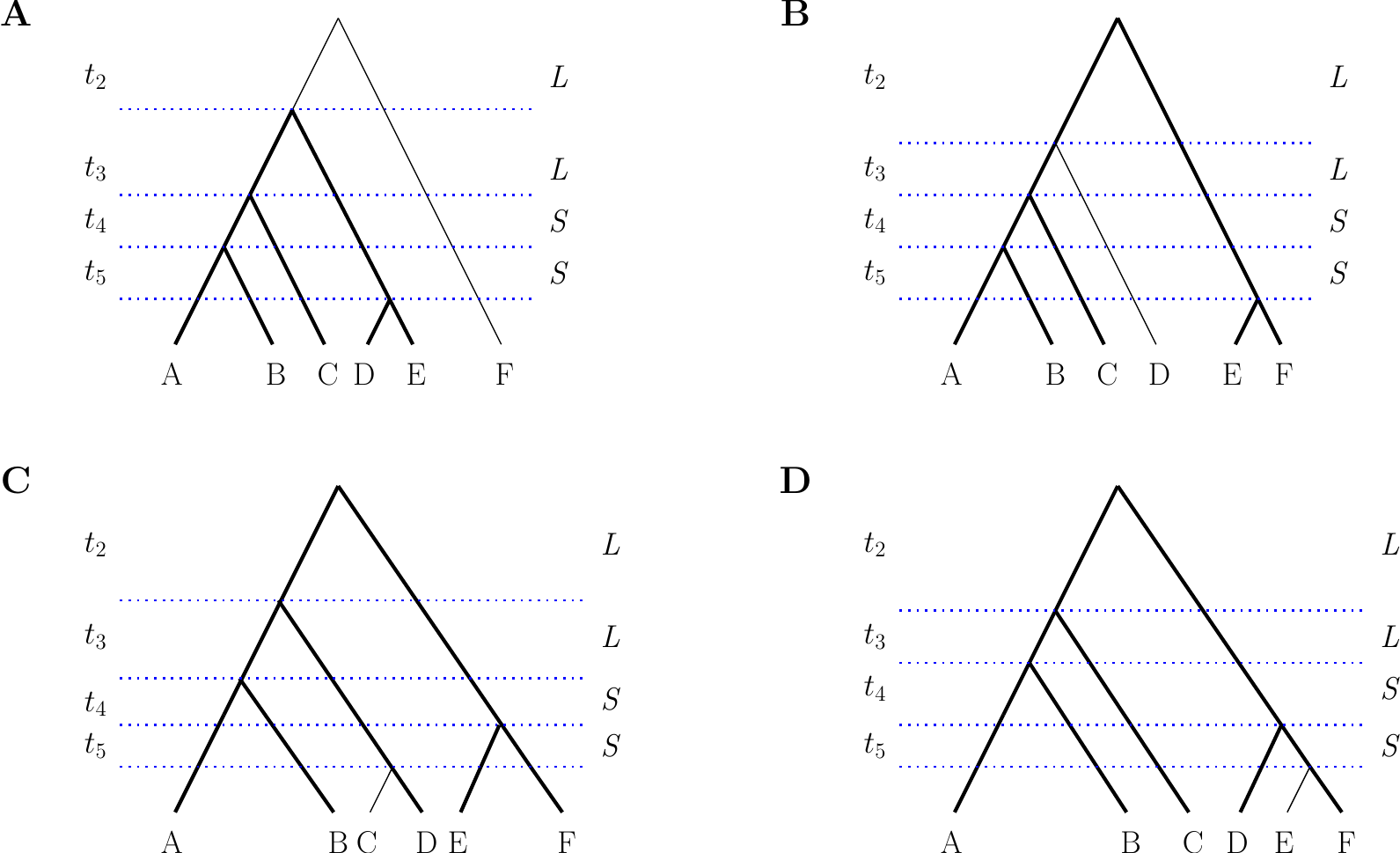}
\caption{Representative labeled rankings of all six-taxon unlabeled species tree topologies, except thecaterpillar and pseudocaterpillar. Bold lines indicate a displayed five-taxon tree topology given in fig.~\ref{fig:5taxa}A. We set some lengths of the speciation intervals to be equal to aid in visualization and computation. Two values $L$ and $S$, measured in coalescent units, are used as interval lengths.  The figures are not drawn to scale. All values of $L$ are equal to each other and all values of $S$ are equal to each other.} 
\label{fig:6taxontops2}
\end{figure}

\begin{figure}[H]
\centering
\includegraphics[width=4in,angle=270]{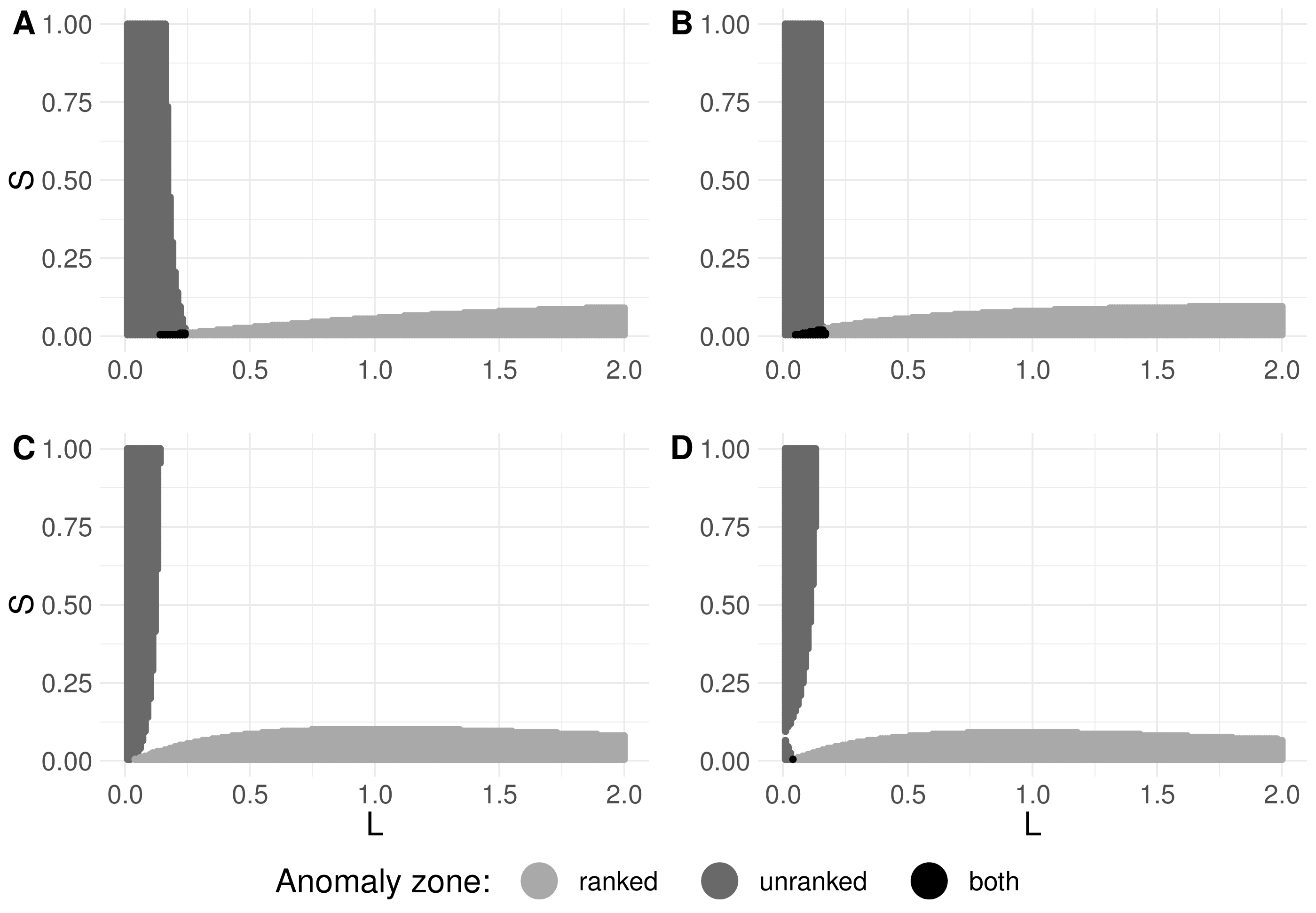}  
\caption{Two-dimensional cross-sections of unranked and ranked anomaly zones, each associated with a six-taxon species tree topology in the corresponding panel of figure~\ref{fig:6taxontops2}. For each species tree topology, 200 values of $L \in [0.01,2]$ and 200 values of $S \in [0.005,1]$ were used to identify the existence of anomalous gene trees.} 
\label{fig:6taxonzones}
\end{figure}

\begin{figure}[H]
\centering
\includegraphics[width=5in]{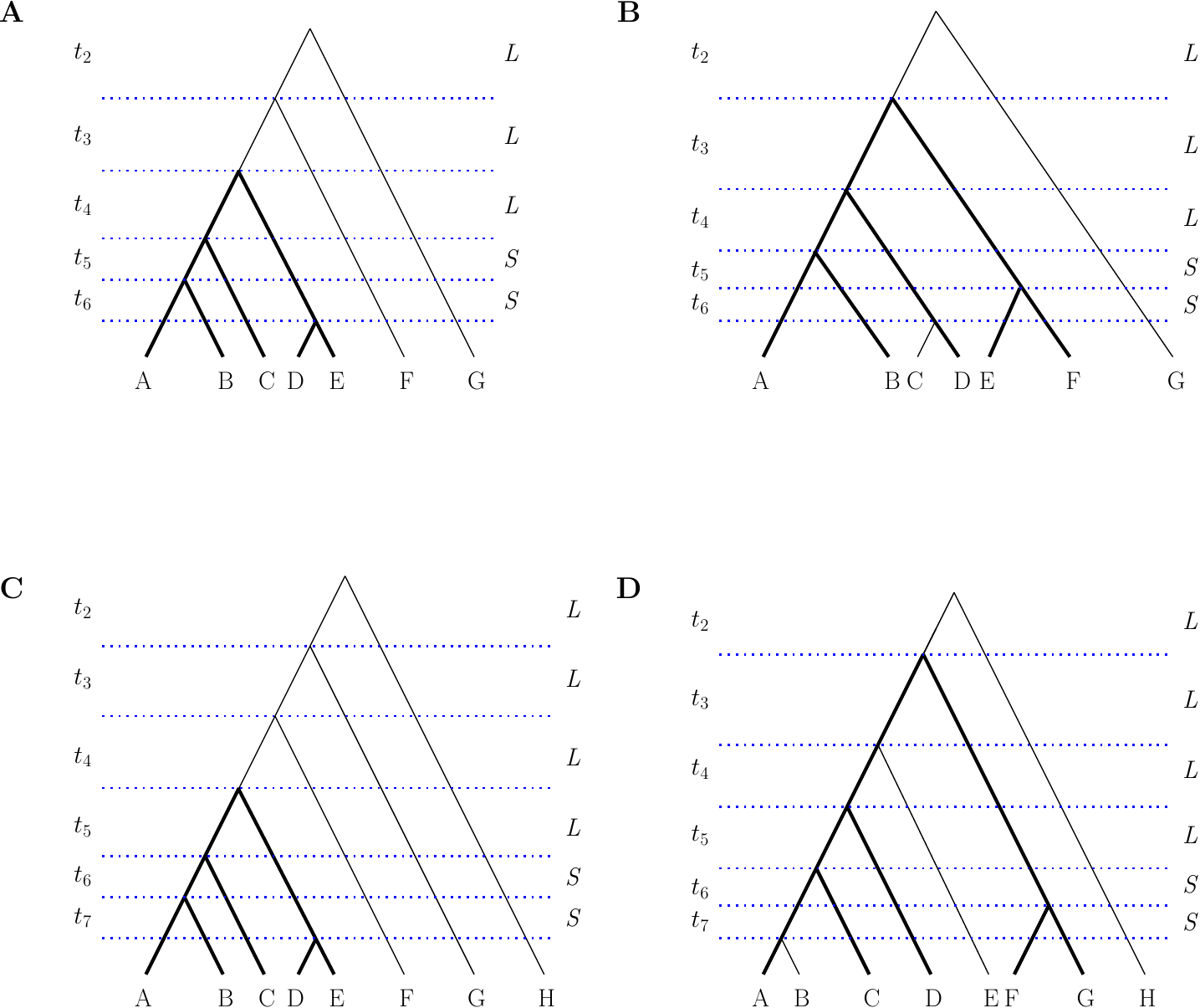}
\caption{Representative labeled rankings of two seven-taxon (top) and two eight-taxon (bottom) species tree topologies that produce anomalous gene trees. Bold lines indicate a displayed five-taxon tree topology given in fig.~\ref{fig:5taxa}A. Two values $L$ and $S$, measured in coalescent units are used as interval lengths. We set some lengths of the speciation intervals to be equal to aid in visualization and computation. Two values $L$ and $S$, measured in coalescent units, are used as interval lengths. The figures are not drawn to scale. All values of $L$ are equal to each other and all values of $S$ are equal to each other.} 
\label{fig:8taxontops}
\end{figure}

Here we use ``caterpillarization" to make seven- and eight-taxon trees resemble the five-taxon ranked tree $(((A,B)_3,C)_2),(D,E)_4)$, the only five-taxon ranked species tree that produces ARGTs. In particular, we consider cases in which a five-taxon species tree topology in figure~\ref{fig:5taxa}A is contained inside the larger trees. This five-taxon tree appears with bold font in larger tree topologies (figures~\ref{fig:6taxontops2} and~\ref{fig:8taxontops}). Because the five-taxon tree in figure~\ref{fig:5taxa}A produces both AGTs and ARGTs, there exists a subset of branch lengths that makes larger trees also have AGTs and ARGTs simultaneously. 

\begin{figure}[H]
\centering
\includegraphics[width=4in,angle=270]{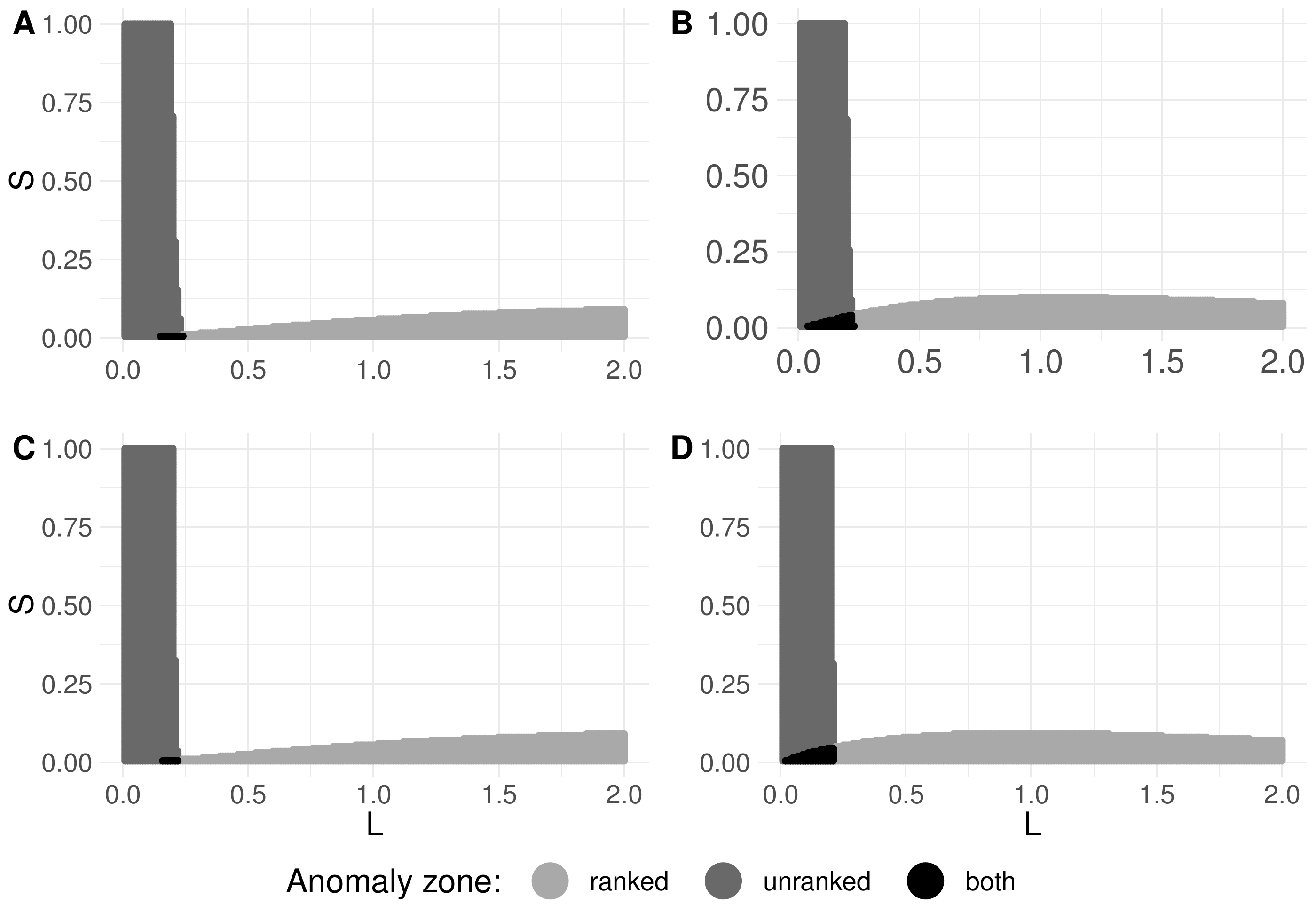}  
\caption{Two-dimensional cross-sections of unranked and ranked anomaly zones for associated seven- and eight-taxon species tree topologies in figure~\ref{fig:8taxontops}. For each species tree topology, 200 values of $L \in [0.01,2]$ and 200 values of $S \in [0.005,1]$ were used to identify the existence of anomalous gene trees.} 
\label{fig:8taxonzones}
\end{figure}

We observe a similar pattern in anomaly zones (fig.~\ref{fig:8taxonzones}) for species tree topologies displayed in figures~\ref{fig:6taxontops2}A,~\ref{fig:8taxontops}A, and~\ref{fig:8taxontops}C. Each of these topologies was obtained from the five-taxon topology in fig.~\ref{fig:5taxa}A by sequentially attaching an additional branch to the root. Under the restriction that speciation intervals have one of two lengths, $S$ and $L$, anomaly zones behave somewhat similar in the cases of $n=6,7,$ and $8$. In particular, the species tree usually needs to have large values of $L$ and small values of $S$ to be in the ranked anomaly zone. However, the pattern is reversed for AGTs: to produce AGTs, $L$ usually needs to be small while $S$ may be relatively large. 

\par
\subsection*{Simulation results}
\par
Next, to explore the probability that random species trees have AGTs and ARGTs, we performed simulations under a birth-death model. In particular, we simulated 5000 species trees with $n=5,6,7,$ and $8$-taxa under a constant rate birth-death model using the \textit{TreeSim} package in R~\citep{Stadler2009, Stadler2011}. In this model, each species at each point in time has the same constant speciation (birth) rate $\lambda$ and extinction (death) rate $\mu$.

\begin{figure}[H]
\centering
\includegraphics[width=.6\textwidth, angle=270,origin=c]{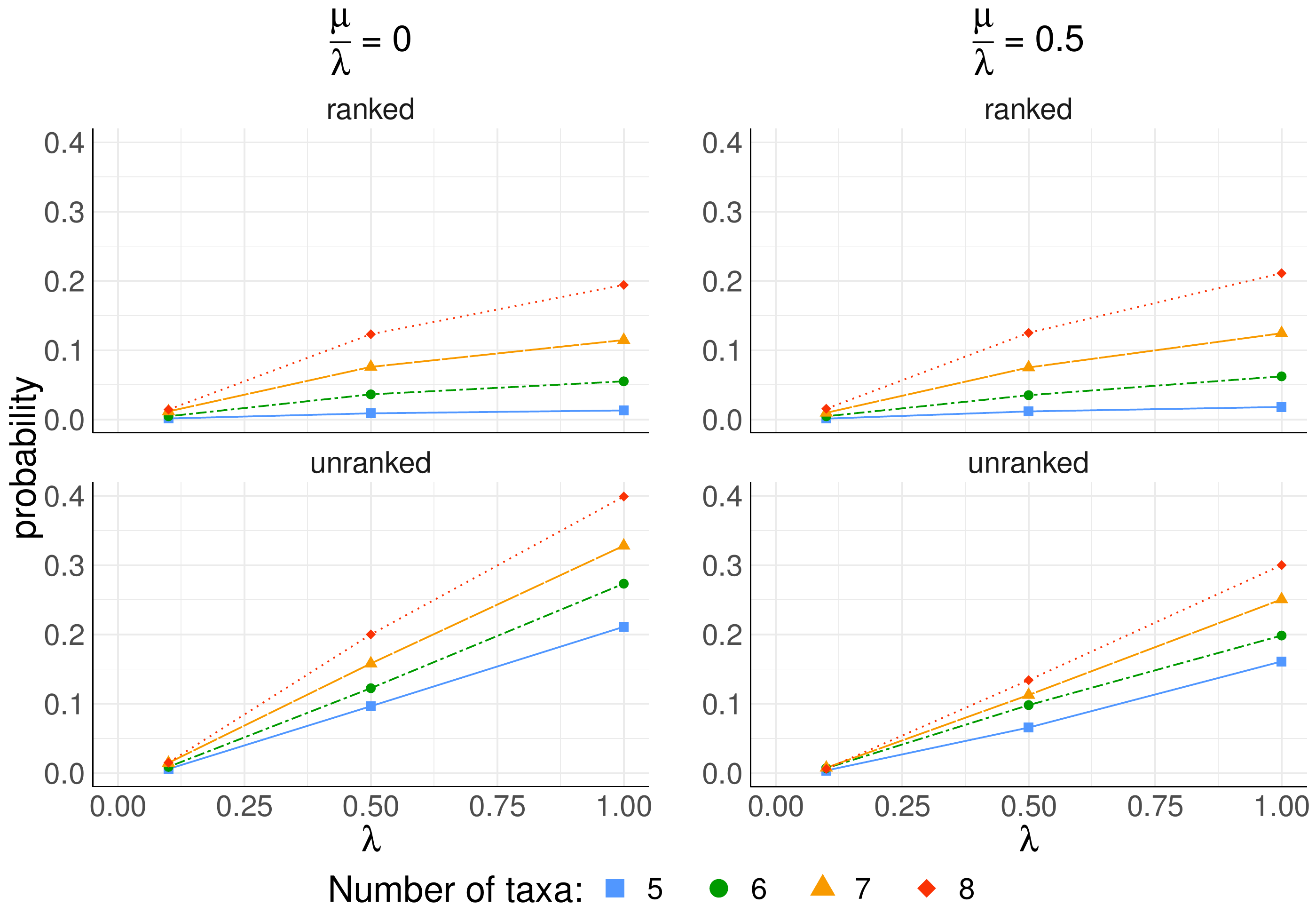}
\caption{The impact of the speciation rate parameter $\lambda$ and the turnover rate $\mu/\lambda$ on the existence of unranked and ranked anomaly zones. For each value of $n = 5, 6, 7,$ and $8$ taxa,  5000 species trees were simulated using a constant rate birth-death process with rates $\lambda=0.1,0.5,$ and $1$ and $\mu/\lambda=0$ and $0.5$. For each combination of $(n, \lambda, \mu)$, the probability of the species tree being in the anomaly zone was computed from the 5000 trials.}
\label{fig:anomprobs}
\end{figure}

Figure~\ref{fig:anomprobs} shows probabilities of the species tree being in the unranked and ranked anomaly zones in relation to the number of taxa $n$, speciation rate $\lambda$, and extinction rate $\mu$. For both types of trees, the probability of a species tree being in an anomaly zone increases with the number of taxa and with $\lambda$. For unranked trees, both results are intuitive: for increasing numbers of taxa, there are more possible ways to have consecutive short branches or intervals in a tree, a pattern typical of the unranked anomaly zone \citep{Rosenberg2013}. Increasing $\lambda$ reduces the average branch length, making consecutive short branches more likely.

We also observed a different effect of the turnover rate $\mu/\lambda$ on the probability of producing unranked and ranked anomalous gene trees. The probability has a decreasing trend for the unranked anomaly zones and an increasing trend for the ranked anomaly zone as turnover rate increases. On average, branch lengths are longer as $\mu$ increases. In particular, a branch length near the root becomes longer, decreasing the probabilities of AGTs but increasing the probabilities of ARGTs.

We calculated the probabilities of ranked and unranked anomaly zones for specific five- and six-taxon tree topologies ($\lambda=0.1,0.5,1$, $\mu=0$, 5000 replicates) to investigate the frequency with which the different tree shapes give rise to AGTs and ARGTs. Under the Yule process, the probabilities of a caterpillar shape, pseudocaterpillar shape, and the unranked version of the tree shape depicted in figure~\ref{fig:5taxa}A for the five-taxon case are 1/3, 1/6, and 1/2. The conditional probabilities of a species tree being in the unranked anomaly zone given the shape are 7.42\%, 0.87\% and 2.15\% for the three shapes, respectively. Because neither caterpillar nor pseudocaterpillar species trees can produce ARGTs, the conditional probabilities of a species tree being in the ranked anomaly zone given the shape are 0\%, 0\% and 0.77\% for the three shapes, respectively.

\begin{figure}[H]
\centering
\includegraphics[width=.7\textwidth,angle=0,origin=c]{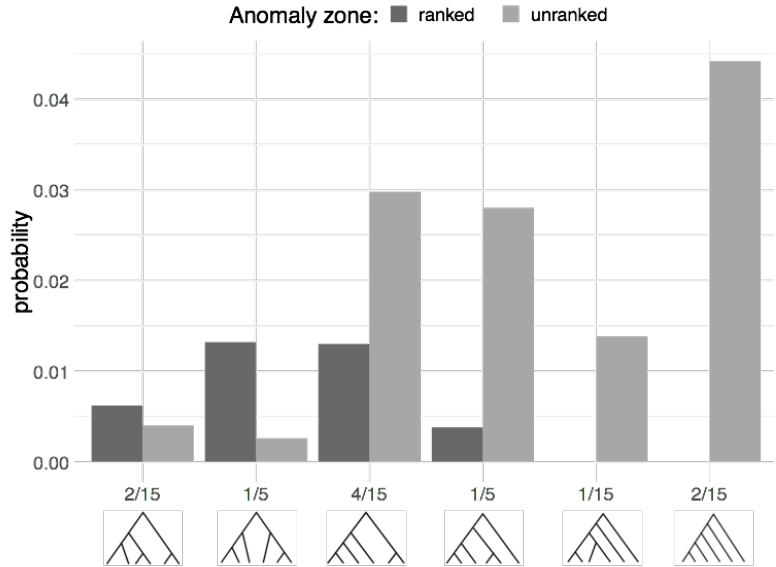}
\caption{Conditional probabilities of ranked and unranked anomaly zones given species tree shape for all possible six-taxon unlabeled, unranked species tree topologies. The exact probabilities of tree shapes under the Yule birth process are displayed on the $x$-axis. The results are based on 5000 species trees simulated under the birth process with $n=6$, $\lambda = 0.5$, and $\mu = 0$. Among the shapes with both AGTs and ARGTs, the third tree shape, with four taxa descended from one side of the root and two from the other, produces the largest combined frequency of AGTs and ARGTs. It is also the most probable shape under the birth process. A similar pattern occurs for $\lambda=0.1$ and $\lambda=1$ (not shown).}
\label{fig:anombar_6taxa}
\end{figure}

Figure~\ref{fig:anombar_6taxa} shows conditional probabilities  of ranked and unranked anomaly zones for all possible six-taxon topologies when $\lambda=0.5$ and $\mu=0$. Under the Yule process the unranked tree shapes have probabilities 2/15, 1/5, 4/15, 1/5, 1/15, and 2/15 from left to right. AGTs arise more often for the caterpillar shape, whereas ARGTs arise more often for the second and third shapes (from left to right). The full probability of anomalous gene trees can be calculated using the law of total probability.

\begin{figure}[H]
\centering
\includegraphics[width=.6\textwidth,angle=270,origin=c]{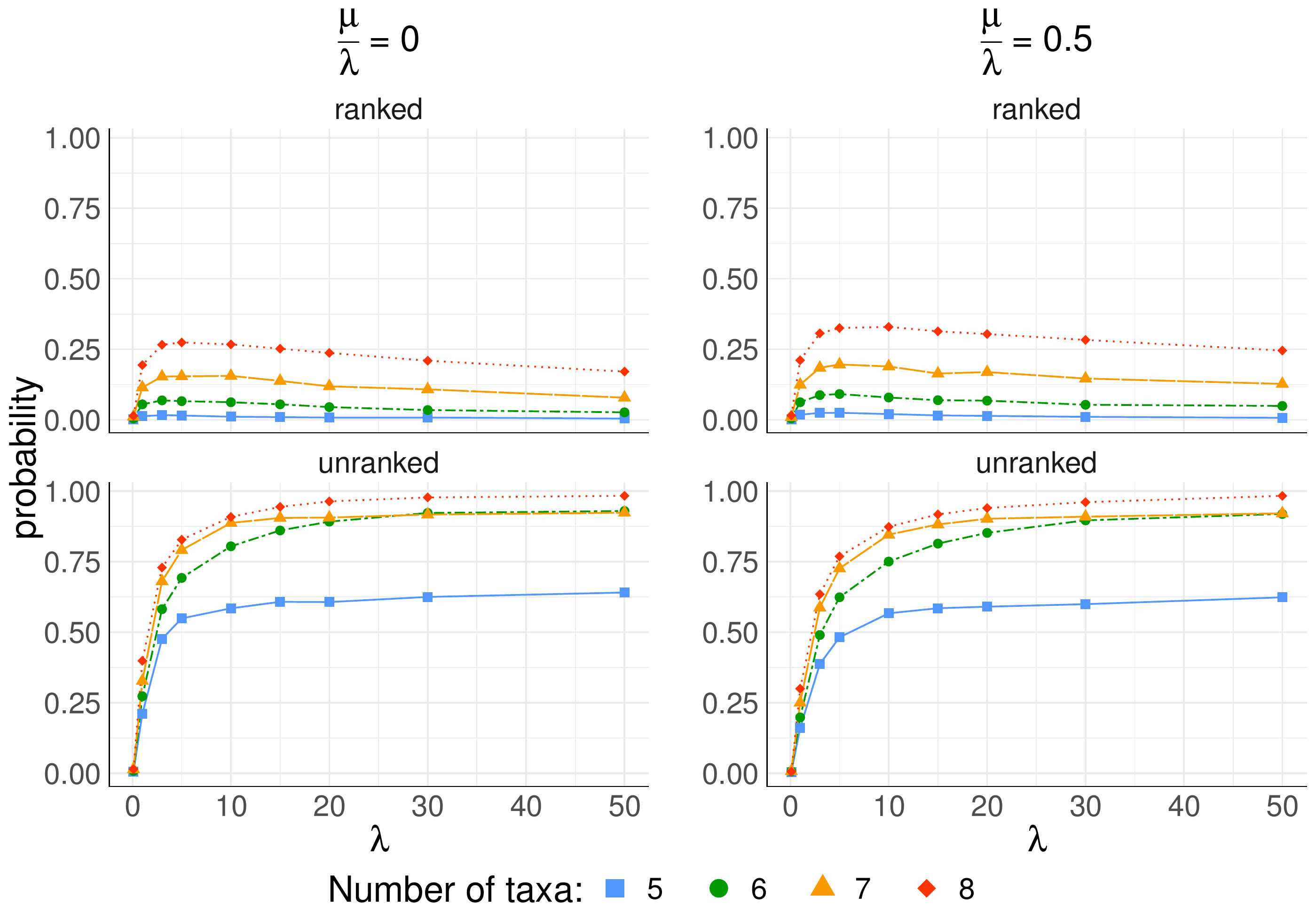}
    \caption{The impact of the speciation rate parameter $\lambda \in [0.1, 50]$ and the turnover rate $\mu/\lambda=0$ and $0.5$ on the existence of unranked and ranked anomaly zones. For each combination of $(n, \lambda, \mu)$, the probability of the species tree being in the anomaly zone was computed from 5000 species trees. Probabilities of the unranked anomaly zone appear to increase with $\lambda$, whereas probabilities of the ranked anomaly zone increase up to a certain value $\lambda \approx 5$, and then begin to decrease.}
\label{fig:anoms_8taxa}
\end{figure}

We also noticed that the probabilities of being in the unranked anomaly zone grow faster than those of the ranked anomaly zone as the speciation rate increases (fig.~\ref{fig:anoms_8taxa}). For example, the probabilities that a species tree belongs to unranked and ranked anomaly zones are equal to $0.399$ and $0.194$, respectively, for $n=8$, $\lambda=1$, and $\mu=0$. For an eight-taxon species tree, with $\lambda=10$ and $\mu=0$, these probabilities are equal to $0.909$ and $0.267$, respectively.

\section*{Discussion}
\par
The existence of anomalous gene trees poses challenges for inferring species trees from gene trees. We have studied AGTs and ARGTs for small trees, identifying cases in which a species tree possesses both types of anomalies (figures~\ref{fig:6taxonzones},~\ref{fig:8taxonzones}). We studied how the parameters of the species tree ($n$, $\lambda$, $\mu$) simulated under a constant rate birth-death process can affect the probability that a species tree is in the anomaly zone. We have shown that often, a species tree has lower probability to be in the ranked anomaly zone than in the unranked anomaly zone (figures~\ref{fig:anomprobs},~\ref{fig:anoms_8taxa}).

We also ran our simulations with larger values of $\lambda$, observing that the probabilities of unranked anomaly zones grow faster than those of ranked anomaly zones as the speciation rate increases (fig.~\ref{fig:anoms_8taxa}). The probability of a species tree being in the ranked anomaly zone for $n=8$ reaches a peak near 27.4\% and begins to decrease for approximately $\lambda > 5$. Probabilities of a species tree being in the unranked anomaly zone appear to increase with $\lambda$, but they are not approaching 1.

An intuitive reason that probabilities do not approach 1 for fixed $n$ is that as $\lambda$ increases, the probability increases that all coalescences occur more anciently than the root of the tree. This scenario does not always result in anomaly zones. For ranked trees, if the species tree is either a caterpillar or pseudocaterpillar, then there cannot be an ARGT, putting a limit on the probability that the species tree lies in the ranked anomaly zone when $n$ is fixed.
In the five-taxon case, ARGTs are more likely when interval $\tau_2$, in which there are two populations (fig.~\ref{fig:genetrees}A), is relatively large compared to other intervals. Increasing $\lambda$ makes this condition less likely.
For unranked species trees, if all coalescences occur above the root, then the species tree has AGTs if, and only if, the species tree does not have a maximally probable shape, where a maximally probable shape is one for which labeled topologies have the maximum number of possible rankings \citep{DegnanAndRosenberg06}. For example, for five taxa, the tree $(((A,B),C),(D,E))$ has three rankings. Thus, if the species tree has this topology and all internal branches have length 0, then no other gene tree shape can be anomalous for it. In this case, as $\lambda \rightarrow \infty$, all unranked labeled gene tree topologies approach probability $r/180$, where $r$ is the number of rankings for the gene tree.

For six taxa, the unlabeled tree shape whose labeled topologies have the maximum number of rankings has four taxa descended from one side of the root and two from the other side, as shown in figure~\ref{fig:6taxontops2}C, where the rooted subtrees on each side of the root themselves maximize the number of possible rankings. This scenario results in an unlabeled tree with eight rankings and $45$ ways to label such tree. Because there are $2700$ ranked labeled topologies for $n=6$ taxa, we therefore expect that as $\lambda\rightarrow\infty$, the probability of the species tree being in an unranked anomaly zone is at least $1 - (45 \cdot 8)/2700 = 13/15$. This value occurs because labeled unranked trees with this maximally probable shape are tied in probability for being the most probable when all coalescences occur more anciently than the root; as $\lambda \rightarrow \infty$, the probability approaches $13/15$ that the species tree does not have the maximally probable shape, and therefore is in an unranked anomaly zone.

More generally, let $T_n$ be an unlabeled species tree shape with the maximum number of rankings. For large $\lambda$, the probability of the species tree with $n$ leaves being in an unranked anomaly zone has a lower bound of   
\begin{align}\label{E:lower_bound}
1 - 2^{n-1-\sigma(T_n)}\displaystyle \prod_{i=1}^{n-1} [c_i(T_n) - 1]^{-1}
\end{align}
where $\sigma(T_n)$ is the number of balanced internal vertices of $T_n$ and $c_i(T_n)$ is the number of descendant leaves of interior vertex $i$, including the root as an interior vertex. The lower bound given in eq.~\eqref{E:lower_bound} can be calculated as $1$ minus the probability that the species tree under the Yule process has the shape that produces the largest number of rankings for a fixed labeling. For example, the lower bound for six-taxon species trees can be calculated as $1-2/15=13/15$. This lower bound in eq.~\eqref{E:lower_bound} underestimates the probability of being in an anomaly zone for large $\lambda$ because even labeled species trees with the maximally probable shape can have AGTs for some sets of branch lengths. It can be shown that this lower bound approaches 1 as $n \rightarrow \infty$ (see \textit{Appendix} for details).

In general, probabilities of both AGTs and ARGTs increase with the number of taxa. For example, the probability of an AGT approximately doubles, going from five to eight taxa for both $\lambda=0.5$ and $\lambda=1$ at both levels of turnover (fig.~\ref{fig:anomprobs}). The probability of an ARGT increases by a factor of $10$ to $15$ going from five to eight taxa at $\lambda=0.5$ and $\lambda=1$ at both levels of turnover (fig.~\ref{fig:anomprobs}).

\begin{figure}[H]
\begin{center}
\includegraphics[width=0.8\textwidth]{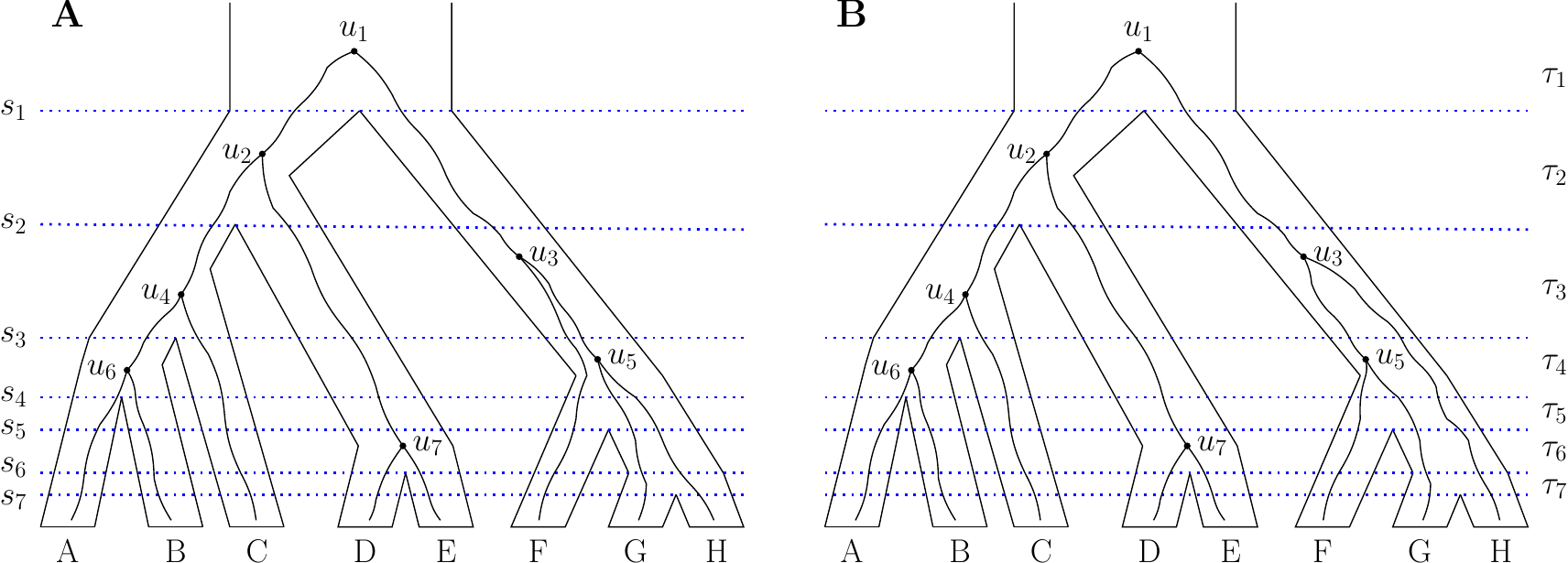}
\caption{Gene trees evolving on an eight-taxon species tree. (A) Ranked gene tree $(((((A,B)_6,C)_4,(D,E)_7)_2,((G,H)_5,F)_3)$ that shares the same unranked topology with that of the species tree. (B) Gene tree $(((((A,B)_6,C)_4,(D,E)_7)_2,((F,G)_5,H)_3)$ that has a different unranked topology from the species tree. Note that the ranked gene tree $(((((A,B)_6,C)_4,(D,E)_7)_2,((F,H)_5,G)_3)$ (not shown) has exactly the same probability as gene trees in (A) and (B) for the species tree depicted. For each $i=1,2,...,7$, $s_i \geq 0$ denotes the time of the $i$th speciation, $\tau_i$ represents the interval between the $(i-1)$th and $i$th speciation events, $t_i$ ($t_i=s_{i-1}-s_i$, $2 \leq i \leq 7$) represents the length of interval $\tau_i$, and $u_i$ represents the $i$th coalescence (node with rank $i$) in the gene tree. The species tree has ranked topology $((((A,B)_4,C)_3,(D,E)_6)_2,((G,H)_7,F)_5)$. For the species tree values $t_i = (0.29,0.006,0.041,0.001,0.022,0.001)$, $i=2,3,...,7$, the ranked gene trees in (A) and (B) are the most probable ranked gene trees, with probability $1.72404 \times 10^{-5}$.} 
\label{fig:tied_probs}
\end{center}
\end{figure}

An open question from \citet{Degnan2012} was whether the most probable ARGT could have a different \textit{unranked} topology from that of the species tree. In that paper, examples of ARGTs had different rankings from the species tree but the same unranked topology.
Here, in our simulation with different combinations of values ($n$, $\lambda$, $\mu$), we have not found any cases where the most probable ranked gene tree and the species tree have different unranked topologies. However, we found a few cases where a gene tree within one step by nearest-neighbor interchange --- which has a different unranked topology from the species tree --- has exactly the same ranked histories and probability as the ranked gene tree topology that matches the unranked species tree topology. For example, for a species tree  given in figure~\ref{fig:tied_probs}, the two ranked gene trees in the figure have the same probabilities, because they have exactly same values of $k_{i,j,z}$ and thus, the same values of $\lambda_{i,j}$ (see eq.~\eqref{E:density2} for details). The same result that at least one of the most probable ranked gene tree topologies must have the same unranked topology as the species tree was proved mathematically by \citet{Disanto2019}. This result suggests that the ``democratic vote'' method used for ranked gene trees might be less misleading than in the unranked setting: if one takes the ranked gene tree (or gene trees, allowing for ties) that occurs most frequently in a large enough sample, then its unranked version is predicted to match the species tree, except possibly when another ranked gene tree is tied for being most probable.

\section*{Materials and Methods}
\subsection*{Calculating the probability of a ranked gene tree topology}
\par
\subsubsection*{General formula}
\par
The probability of the ranked gene tree $P(\mathcal{G}|\mathcal{T})$ can be computed as a sum over all ranked histories. Denote the probability in interval $\tau_i$ for a particular ranked history $x$ by $P(\mathcal{G}_{\tau_i},x|T)$. The probability of a ranked gene tree topology $\mathcal{G}$ with ranked history set $\mathcal{Y}$  given a species tree $\mathcal{T}$ can be written
\begin{equation}\label{E:general}
    P\left(\mathcal{G}|\mathcal{T}\right) = \sum_{x \in \mathcal{Y}} H_{\ell_1}(x) \displaystyle \prod_{i=2}^{n-1} P(\mathcal{G}_{\tau_i},x|{\cal T}), 
\end{equation}
    where $H_{\ell_1}(x)$ is that probability for the coalescences above the root appear in the order that follows the ranked gene tree \citep{Stadler2012}. If the number of lineages above the root is $\ell_1$, then \citep{Rosenberg2006}
\begin{equation}\label{E:H}
    H_{\ell_1}(x) = \frac{2^{\ell_1-1}}{\ell_1! (\ell_1-1)!}. 
\end{equation}

Denote the number of lineages available for coalescence in population \textit{z} just after (going forward in time) the $j$th coalescence in interval $\tau_i$ by $k_{i,j,z}$. The probability that $\ell$ lineages fail to coalesce in a time interval of length $t_i$ is $e^{-\binom{\ell}{2} t_i}$. Hence, the waiting time until the next coalescent event (going backward in time) has rate $\lambda_{i,j} = \displaystyle \sum_{z=1}^i \binom{k_{i,j,z}}{2}$. The density for the coalescent events in the interval $\tau_i$ is \citep{Degnan2012}
    \begin{equation}\label{E:density}
    f_i(v_0,v_1,...,v_{m_i}) = \exp\left(- \sum_{j=0}^{m_i} \lambda_{i,j}v_j\right),
    \end{equation}
where $v_j$ is the time between the $j$th and $(j+1)$st coalescent events, with $v_0$ being the time between $s_{i-1}$ and the least recent coalescent event in $\tau_i$ and with $v_{m_i}$ being the time between $s_i$ and coalescent event $m_i$.

For example, consider the second speciation interval $\tau_2$ for the species tree in fig.~\ref{fig:genetrees}A. Here, $v_0$ is the time between $s_1$ and the least recent coalescent event $u_2$ in interval $\tau_2$. Similarly, $v_1$ is the time between $u_2$ and $u_3$, $v_2$ is the time between $u_3$ and $u_4$, and $v_{m_i}=v_3$ is the time between $u_4$ and $s_2$. Using the fact that the sum of exponential random variables with different rates $\lambda_i$ has hypoexponential distribution, eq.~\eqref{E:density} can be written as follows \citep{Stadler2012}: 
\begin{equation}\label{E:density2}
    P\left(\mathcal{G}_{\tau_i}, x|\mathcal{T}\right) = \int_v f_i(v_0,...,v_{m_i}) dv = \sum_{j=0}^{m_i} \frac{e^{-\lambda_{i,j}(s_{i-1}-s_i)}}{\displaystyle\prod_{k=0,k\neq j}^{m_i}(\lambda_{i,k}-\lambda_{i,j})}.
\end{equation}

\par
\subsubsection*{Examples}
\par
Consider a species tree $\mathcal{T}$ and gene tree with matching ranked topology $((A,(B,(C,D)_4)_3)_2,(E,F)_5)$ (fig.~\ref{fig:genetrees}C). We now calculate the probability of the ranked history $(1,2,2,2,2)$ in interval $\tau_2$. Because four coalescences occur in interval $\tau_2$, $m_2=4$ and $k_{2,j,z}$ is defined for $j=0,1,2,3,4$ and $z=1,2$. We have $k_{2,j,1}=(1, 2, 3, 4, 4)$ for $j=0,1,...,4$ and $k_{2,j,2}=(1, 1, 1, 1, 2)$ for $j=0,1,...,4.$
Using $\lambda_{2,j} = \displaystyle \sum_{z=1}^2 \binom{k_{2,j,z}}{2}$, we have $\lambda_{2,j}=(0, 1, 3, 6, 7),$ for $j=0,1,...,4.$
Thus, eq.~\eqref{E:density2}  evaluates to  
\begin{equation}
    P\left(\mathcal{G}_{\tau_2},(1,2,2,2,2)|\mathcal{T}\right) = \sum_{j=0}^{4} \frac{e^{-\lambda_{2,j}t_2}}{\displaystyle\prod_{k=0,k\neq j}^{4}(\lambda_{2,k}-\lambda_{2,j})} = \frac{1}{126} - \frac{e^{-t_2}}{60} + \frac{e^{-3t_2}}{72} - \frac{e^{-6t_2}}{90} + \frac{e^{-7t_2}}{168}, \nonumber
\end{equation}
where $t_2=s_1-s_2$ is the length of interval $\tau_2$.

Similarly, we can compute the probabilities in intervals $\tau_3$, $\tau_4$, $\tau_5$. Given that the probability for the coalescence of $\ell_1 = 2$ lineages above the root appearing in the right order is $H_{2} = 1$ \eqref{E:H},  the probability of the ranked history $(1,2,2,2,2)$ is equal to
\begin{align}
    P\left(\mathcal{G},(1,2,2,2,2)|{\cal T} \right) = &H_{2}(x) \cdot \prod_{i=2}^{5} P\left(\mathcal{G}_{\tau_i},(1,2,2,2,2)|{\cal T}\right) \nonumber \\
    =&\left(\frac{1}{126} - \frac{e^{-t_2}}{60} + \frac{e^{-3t_2}}{72} - \frac{e^{-6t_2}}{90} + \frac{e^{-7t_2}}{168}\right) 
    \cdot e^{-4t_3 - 2t_4 -t_5},\label{eq6}
\end{align}
where $t_i=s_{i-1}-s_i.$

\par
Now consider a species tree $\mathcal{T}$ with nonmatching ranked topology $((A,(B,(C,D)_5)_4)_2,(E,F)_3)$ (fig.~\ref{fig:genetrees}D). The values of $k_{i,j,z}$ in interval $\tau_2$ are

$k_{2,j,1}=(1, 2, 2, 3, 4),$ $j=0,1,...,4;$
$k_{2,j,2}=(1, 1, 2, 2, 2),$ $j=0,1,...,4.$

Thus, $\lambda_{2,j}=(0, 1, 2, 4, 7)$ for $j=0,1,...,4$, and the probability of the nonmatching ranked gene tree for the ranked history $(1,2,2,2,2)$ is
\begin{align}
    P\left(\mathcal{G},(1,2,2,2,2)|{\cal T} \right) = &H_{2}(x) \cdot \prod_{i=2}^{5} P\left(\mathcal{G}_{\tau_i},(1,2,2,2,2)|{\cal T}\right) \nonumber \\
    =& \left(\frac{1}{56} - \frac{e^{-t_2}}{18} + \frac{e^{-2t_2}}{20} - \frac{e^{-4t_2}}{72} + \frac{e^{-7t_2}}{630}\right) \cdot e^{-4t_3-2t_4-t_5}. \label{eq7}
\end{align}
Following eqs.~(\ref{eq6}) and~(\ref{eq7}), the limiting probabilities for the matching and nonmatching ranked gene tree topologies for the ranked history $(1, 2, 2, 2, 2)$ when $t_2 \rightarrow \infty$ and $t_3,t_4,t_5 \rightarrow 0$ are $\frac{1}{126}$ and $\frac{1}{56}$ respectively. Thus, the ranked history $(1, 2, 2, 2, 2)$ is more probable for the nonmatching ranked gene tree topology than for the matching ranked history when $t_2 \rightarrow \infty$ and $t_3,t_4,t_5 \rightarrow 0$. For sufficiently large $t_2$ and sufficiently small $t_3,t_4,t_5$, most of the probability of the ranked gene tree topology is concentrated on this ranked history, making the probabilities of the other ranked histories close to 0. Thus, the most probable ranked gene tree topology becomes discordant from the ranked species tree topology, forcing the species tree into the ranked anomaly zone.
\par

 http
\subsubsection*{\textit{PRANC} software}
\par
 We implemented the program \textit{PRANC}, which can analytically compute the probabilities of ranked gene trees given a species tree in Newick format, following eq.~\eqref{E:general}. The program has an option to compute the probability of an unranked gene tree by summing the probabilities of all ranked gene trees that share the corresponding unranked topology. We improved the numerical results by adding the probabilities of the ranked histories in ascending order, enabling the smallest-magnitude values to accumulate before interacting with larger-magnitude values. In addition, \textit{PRANC} has an option to output symbolic probabilities followed by ranked histories ({\tt https://github.com/anastasiiakim/PRANC}).
\begin{verbatim}
pranc -rprob <species-tree-file-name> <ranked-gene-tree-file-name>
pranc -uprob <species-tree-file-name> <unranked-gene-tree-file-name>
pranc -sym <species-tree-file-name> <ranked-gene-tree-file-name>
\end{verbatim} 

\textit{PRANC} also can output the ``democratic vote'' ranked or unranked tree topology, respectively. The program will output two files: one with ranked/unranked topologies for each tree, and another with unique topologies and their frequencies. 
\begin{verbatim}
pranc -rtopo <input-file-name>
pranc -utopo <input-file-name>
\end{verbatim}

\subsection*{Simulations}

We simulated species phylogenies under a constant rate birth-death model. In this model, each species is equally likely to be the next to speciate. Each tree branch gives birth to a new branch at rate $\lambda$. Lineages can also go extinct at rate $\mu$.

Because the length of a randomly selected interior branch in a Yule (rate $\lambda$) tree on $n$ leaves is exponentially distributed with rate $2\lambda$ \citep{StadlerandSteel}, for $\lambda=0.1$ and $\lambda=1$ a species tree has a mean branch length of $1/(2\cdot0.1)=5$ and $1/(2\cdot1)=0.5$ respectively. We note that if all branch lengths were 0.5 coalescent units, then the
species trees in the simulations would be outside of the unranked anomaly zone.  A value of 0.5 coalescent units
for an internal branch means that two lineages have a probability of coalescing of $1-\exp(-0.5) \approx 39\%$ of coalescing within that branch, whereas for 5 coalescent units, the probability of coalescence exceeds 99\%. Values of $\lambda$ near 0.5 are chosen to be reasonably plausible for hominid evolution \citep{TreeBalance2016}. The range of $\lambda = 0.1$ to $\lambda = 1$ thus gives a range of low to moderate levels of incomplete lineage sorting that are plausibly consistent with empirical studies.

We let the speciation rate $\lambda$ take the values of $0.1,0.5,$ and $1$, and choose the extinction rate $\mu$ to depend on $\lambda$ such that the turnover rate $\mu/\lambda$ is $0$ or $0.5$. Values of $(n, \lambda,\mu)$ were chosen to examine the effect of the species tree parameters on the existence of anomalous gene trees. For each combination $(n, \lambda,\mu)$, the distributions of unranked and ranked gene tree topologies were computed analytically for each simulated species tree. The probabilities of all possible unranked and ranked topologies were computed using \textit{hybrid-coal} \citep{Zhu2017} and \textit{PRANC} respectively, conditional on a species tree generated under a constant rate birth-death model with parameters $(n, \lambda,\mu)$. The presence of anomalous gene trees was then identified by comparing the analytical probabilities of the matching gene tree topology and the most probable nonmatching gene tree topology.

\par
\section*{Acknowledgments}
\par
This work was supported by National Institute of Health R01 grants GM117590 and GM131404.

\bibliographystyle{plainnat}
\bibliography{rankedbib}

\newpage
\section*{Appendix}
Here we prove the lower bound in eq.~\eqref{E:lower_bound} of the probability of the species tree with $n$ leaves being in an unranked anomaly zone for large $\lambda$, and we show that this lower bound approaches $1$ as $n \rightarrow \infty$ and $\lambda \rightarrow \infty$.

Let $T_n$ be a labeled species tree whose unlabeled shape maximizes the number of rankings. of its associated labeled topologies. For large $\lambda$, the probability of the species tree with $n$ leaves being in an unranked anomaly zone has a lower bound of 
\begin{align}\label{E:lower_bound_full}
1 - \frac{N_R \cdot R}{N_T},
\end{align}
where $N_R$ is number of ways to label the unranked unlabeled tree with the maximum number of rankings, $R$ is the number of rankings, and $N_T$ is the number of ranked topologies for an $n$-taxon labeled tree. 

A given unlabeled tree topology has $R=(n-1)!/\prod_{i=1}^{n-1} (c_i-1)$ rankings, where $c_i$ is the number of descendant leaves of interior vertex $i$, including the root as an interior vertex \citep[][p.~46]{steel2016phylogeny}. There are $N_R=n!2^{-\sigma}$ ways to label the tree with the maximum number of rankings, where $\sigma$ is the number of balanced internal vertices \citep{steel2016phylogeny}. Because the number of ranked topologies for an $n$-taxon tree is $N_T=\prod_{i=2}^n {i \choose 2} = n!(n-1)!/2^{n-1}$ \citep{brown1994,steel2016phylogeny}, equation~\eqref{E:lower_bound_full} leads to the following expression:
\begin{align}\label{E:lower_bound_recall}
1 - \frac{n!2^{-\sigma(T_n)} \cdot (n-1)! \prod_{i=1}^{n-1} [c_i(T_n)-1]^{-1} }{n!(n-1)!/2^{n-1}} = 1 - \frac{2^{n-1-\sigma(T_n)}}{(n-1) \prod_{i=2}^{n-1} [c_i(T_n) - 1]},
\end{align}
equivalent to the expression \eqref{E:lower_bound}.

\begin{figure}[H]
\centering
\includegraphics[width=4in,angle=270]{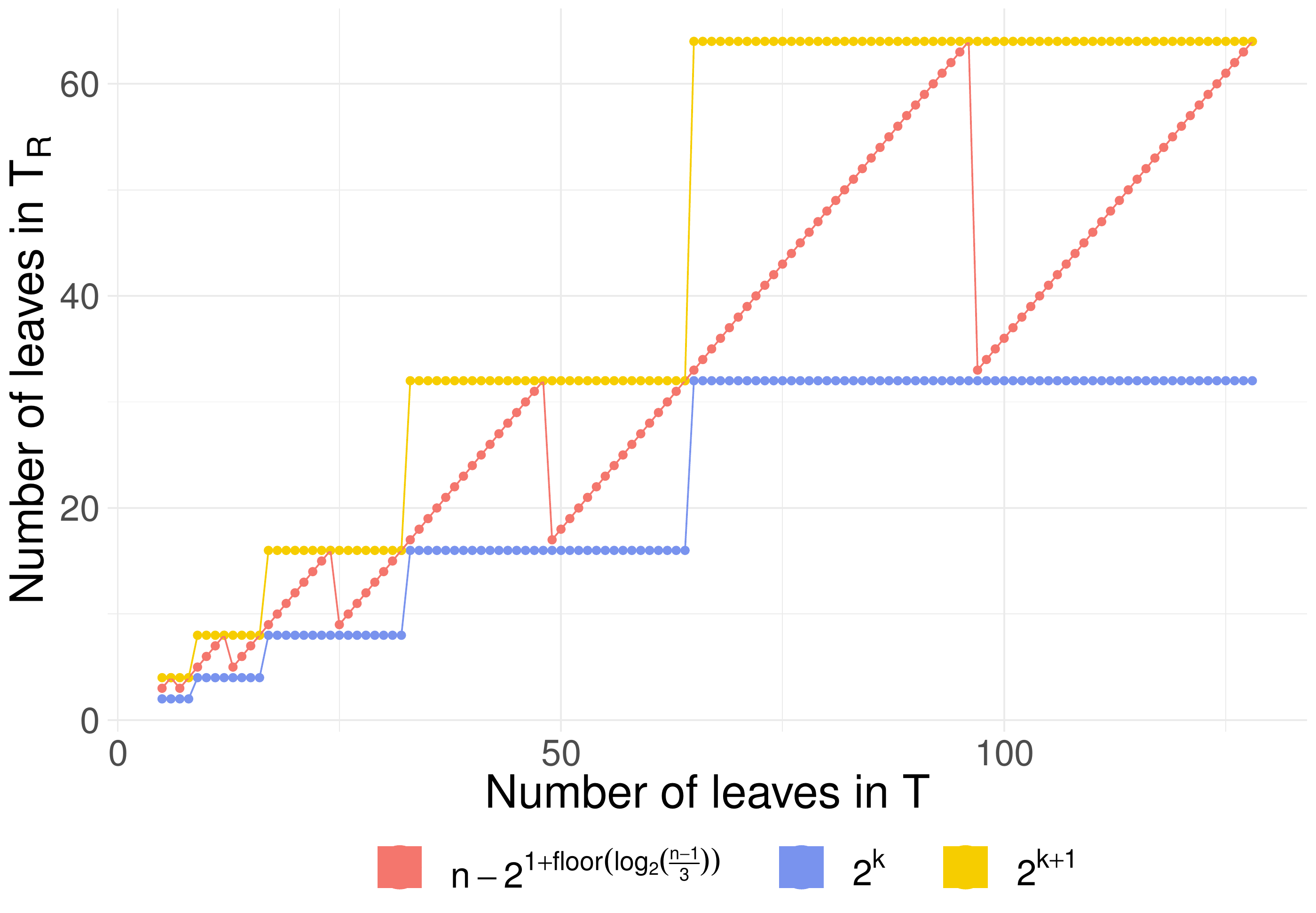}  
\caption{The values of $n-2^{1 + \floor*{\log_2[(n-1)/3]}}$, $2^k$, and $2^{k+1}$ for a tree with $2^{k+1} < n \leq 2^{k+2}$ taxa. The tree with the maximum number of rankings has $2^k < 2^{1 + \floor*{\log_2[(n-1)/3]}} \leq 2^{k+1}$ taxa descended from one side of the root and $2^k < n-2^{1 + \floor*{\log_2[(n-1)/3]}} \leq 2^{k+1}$ from the other side.} 
\label{fig:appendix}
\end{figure}

An $n$-taxon labeled species tree $T_n$ with the maximum number of rankings has $2^{1 + \floor*{\log_2[(n-1)/3]}}$ taxa descended from one side of the root and $n-2^{1 + \floor*{\log_2[(n-1)/3]}}$ from the other side \citep{Harding1971,Harding1974,Hammersley1974} (table~\ref{table:leftright}). For an $n$-taxon tree, $n$ must be between two powers of 2.  
Let $k$ be an integer with $2^{k+1} < n \leq 2^{k+2}$. For a tree with the maximum number of rankings, one of the subtrees descended from $T_n$ has at most $2^{k+2}$ leaves and has the number of leaves a power of $2$, the tree should have at most $2^{k+1}$ leaves. In particular, $T_n$ with $2^{k+1} < n \leq 2^{k+2}$ leaves has $2^k < 2^{1 + \floor*{\log_2[(n-1)/3]}} \leq 2^{k+1}$ taxa descended from one side of the root and $2^k < n-2^{1 + \floor*{\log_2[(n-1)/3]}} \leq 2^{k+1}$ from the other side (table~\ref{table:leftright}, figure~\ref{fig:appendix}). The tree rooted on each side of the root of $T_n$ itself maximizes the number of possible rankings for all labeled trees with the same number of leaves.

 To prove that the lower bound approaches $1$ as $n \rightarrow \infty$, we need to show that in eq.~\eqref{E:lower_bound_recall},  $\prod_{i=2}^{n-1} [c_i(T_n) - 1]^{-1} \rightarrow 0$ and $2^{n-1-\sigma(T_n)} (n-1)^{-1} \leq 1$ as $n \rightarrow \infty$. We consider three cases: (1) $n=2^{k+2}$, (2) $n$ odd, and (3) $n$ even and $n \neq 2^{k+2}$.

Consider a case with $n=2^{k+2}$, $k=0,1,...$ . A completely balanced symmetric shape is the shape with the maximum number of rankings, with $\sigma(T_n)=n-1$. Thus, for $n=2^{k+2}$, eq.~\eqref{E:lower_bound_recall} can be written as follows:
\begin{align}\label{E:even2k}
1 - \displaystyle \prod_{i=1}^{k+1}(2^{k-i+3} - 1)^{-2^{i-1}}.
\end{align}
The product in eq.~\eqref{E:even2k} is the inverse product of the numbers of descendant leaves of all interior vertices, including the root as an interior vertex. That the lower bound for $n=2^{k+2}$ approaches $1$ as $k \rightarrow \infty$ (see Lemma 1 for proof) is proven by

\textbf{Lemma 1}: Let $c_i(T_n)$ be the number of descendant leaves of interior vertex $i$ of a tree $T_n$, excluding the root. Then $\prod_{i=2}^{n-1} [c_i(T_n) - 1]^{-1} \rightarrow 0$ as $n \rightarrow \infty$.
\begin{proof}
Define $c_i^*$ as
\begin{equation}\nonumber
  c_i^*=\begin{cases}
    2, & \text{if $i$ is a cherry},\\
    3, & \text{otherwise}.
  \end{cases}
\end{equation}

The maximum number of cherries of an $n$-taxon tree is at most $n/2$. Hence,
\begin{align}\nonumber
\prod_{i=2}^{n-1} [c_i(T_n)-1]^{-1} \leq \prod_{i=2}^{n-1} [c_i^*(T_n)-1]^{-1} \leq 2^{-(n-2-n/2)} = 2^{-n/2+2},
\end{align}
where $n-2-n/2$ is the number of internal nodes excluding the root minus the maximum number of cherries. This quantity approaches $0$ as $n \rightarrow \infty$, completing the proof. $\qedhere$ 
\end{proof}

For the other two cases, we use a series of lemmas.

\textbf{Lemma 2}: Let $\sigma(T_n)$ be the number of balanced internal vertices in $T_n$, the tree with the maximal number of rankings. Then $\sigma(T_n) = n-k-1$ when $n$ is odd and $2^{k}<n<2^{k+1}$.

\begin{proof}
Let $C(k)$ be the statement that for odd $n$ and $2^{k}<n<2^{k+1}$, $\sigma(T_n)=n-k-1$. $C(k)$ is true for $k=1$ since $3$-taxon trees have one balanced internal vertex. Now we show that if $C(k)$ is true, then $C(k+1)$ is true for any $k \geq 1$.

We need to show that for odd $n$, $2^{k+1} < n < 2^{k+2}$, the number of balanced internal vertices is $\sigma(T_n)=n-(k+1)-1=n-k-2$.

Among trees with $2^{k+1}<n<2^{k+2}$ leaves, let $T_n$ be a tree with the maximal number of rankings. Let $\ell(T_L)$ and $\ell(T_R)$ be the numbers of leaves in the trees rooted at the left and right immediate descendants of the root respectively. Without loss of generality, let $\ell(T_L)=2^{1 + \floor*{\log_2[(n-1)/3]}}$ and $\ell(T_R)=n-2^{1 + \floor*{\log_2[(n-1)/3]}}$.

$T_L$ is a completely balanced symmetric tree, $\sigma(T_L)=2^{1 + \floor*{\log_2[(n-1)/3]}}-1$. Because $n$ is odd, $T_R$ has an odd number of leaves with $2^k < n-2^{1 + \floor*{\log_2[(n-1)/3]}} < 2^{k+1}$ for $2^{k+1} < n < 2^{k+2}$ (figure~\ref{fig:appendix}).

Now, using an induction assumption that $C(k)$ is true,
$\sigma(T_n) = \sigma(T_L)+\sigma(T_R) = 2^{1 + \floor*{\log_2[(n-1)/3]}} - 1 + ( n-2^{1 + \floor*{\log_2[(n-1)/3]}} - k - 1) = n - k - 2$.
$\qedhere$ 
\end{proof}

\textbf{Lemma 3}: Let $\sigma(T_n)$ be the number of balanced internal vertices in $T_n$, the tree with the maximal number of rankings. Then $\sigma(T_n) \geq n-k-1$ when $n$ is even and $2^{k}<n\leq2^{k+1}$, $k \ge 0$.

\begin{proof}
Let $C(k)$ be the statement that for even $n$ and $2^{k}<n\leq2^{k+1}$, $\sigma(T_n) \geq n-k-1$. Obviously, $C(k)$ is true for $k=0$ since $2$-taxon trees have one balanced internal vertex ($\sigma(T_2) \geq 1$). Now we show that if $C(k)$ is true, then $C(k+1)$ is true for any $k \geq 0$.

We need to show that for even $n$, $2^{k+1} < n \leq 2^{k+2}$, the number of balanced internal vertices is $\sigma(T_n) \geq n-(k+1)-1=n-k-2$.

Among trees with $2^{k+1}<n\leq2^{k+2}$ leaves, let $T_n$ be a tree with the maximal number of rankings. Let $\ell(T_L)$ and $\ell(T_R)$ be the numbers of leaves in the trees rooted at the left and right immediate descendants of the root respectively. Without loss of generality, let $\ell(T_L)=2^{1 + \floor*{\log_2[(n-1)/3]}}$ and $\ell(T_R)=n-2^{1 + \floor*{\log_2[(n-1)/3]}}$.

$T_L$ is a completely balanced symmetric tree, $\sigma(T_L)=2^{1 + \floor*{\log_2[(n-1)/3]}}-1$. Because $n$ is even, $T_R$ has an even number of leaves with $2^k < n-2^{1 + \floor*{\log_2[(n-1)/3]}} \leq 2^{k+1}$ for $2^{k+1} < n \leq 2^{k+2}$ (figure~\ref{fig:appendix}).

Now, using an induction assumption that $C(k)$ is true,
$\sigma(T_n) = \sigma(T_L)+\sigma(T_R) \geq 2^{1 + \floor*{\log_2[(n-1)/3]}} - 1 + ( n-2^{1 + \floor*{\log_2[(n-1)/3]}} - k - 1) = n - k - 2$.
$\qedhere$ 
\end{proof}

\textbf{Lemma 4}: $2^{n-1-\sigma(T_n)} (n-1)^{-1} \leq 1$ as $n \rightarrow \infty$.

\begin{proof}
From Lemmas 2 and 3, it follows that $\sigma(T_n) \geq n-k-1$ for $2^k < n \leq2^{k+1}$ and $\log_2(n) - 1 \leq k < \log_2(n)$. 

Consider two cases: $k=\log_2(n)-1$ and $\log_2(n) - 1 < k < \log_2(n)$. If $k=\log_2(n)-1$, then $\sigma(T_n) \geq n-\log_2(n)$ and
\begin{align}\nonumber
2^{n-1-\sigma(T_n)} \leq 2^{\log_2(n)-1} =  2^{\log_2(n)}/2 = n/2 \leq n-1.
\end{align}

From $\log_2(n) - 1 < k < \log_2(n)$ and the fact that $k$ is an integer, $k=\floor{\log_2(n)}$ and $\sigma(T_n) \geq n-1-\floor{\log_2(n)}$. Then, as $n \rightarrow \infty$
\begin{align}\nonumber
2^{n-1-\sigma(T_n)} \leq 2^{\floor{\log_2(n)}} \leq 2^{\log_2(n-1)} = n-1.
\end{align}

It follows that, as $n \rightarrow \infty$,
\begin{align}\nonumber
2^{n-1-\sigma(T_n)} (n-1)^{-1} \leq (n-1)/(n-1) = 1. 
\end{align} $\qedhere$
\end{proof}

\textbf{Theorem}: The lower bound of the probability of the species tree with $n$ leaves being in an unranked anomaly zone, as defined in eq.~\eqref{E:lower_bound_recall}, approaches $1$ as $n \rightarrow \infty$ and $\lambda \rightarrow \infty$.

\begin{proof}
The result immediately follows by Lemmas 1 and 4 in eq.~\eqref{E:lower_bound_recall}.
$\qedhere$ 
\end{proof}

\par
\begin{table}[!htbp]
\begin{center}
\caption{The $n$-taxon species trees with the maximum number of rankings for a labeled topology.}
\begin{tabular}{r r r r  r r r r  r r r r  r r r r  r}
\cmidrule[0.5pt]{1-17}
\multicolumn{1}{c}{$n$}&\multicolumn{1}{c}{$(\ell,r)$}&\multicolumn{1}{c}{}&\multicolumn{1}{c}{}&\multicolumn{1}{c}{}&\multicolumn{1}{c}{$n$}&\multicolumn{1}{c}{$(\ell,r)$}&\multicolumn{1}{c}{}&\multicolumn{1}{c}{}&\multicolumn{1}{c}{}&\multicolumn{1}{c}{$n$}&\multicolumn{1}{c}{$(\ell,r)$}&\multicolumn{1}{c}{}&\multicolumn{1}{c}{}&\multicolumn{1}{c}{}&\multicolumn{1}{c}{$n$}&\multicolumn{1}{c}{$(\ell,r)$}\\\midrule
\multicolumn{1}{ c }{2} & (1,1) & & & & 18 & (10,8) & & & & 34 & (18,16) & & & & 50 & (32,18) \\ 
\multicolumn{1}{ c }{3} & (2,1) & & &  & 19  & (11,8)  & & & & 35 & (19,16) & & & & 51 & (32,19) \\ 
\multicolumn{1}{ c }{4} & (2,2) & & &  &  20 & (12,8)  & & & & 36 & (20,16) & & & & 52 & (32,20) \\ 
\multicolumn{1}{ c }{5} & (3,2) & & &  &  21 & (13,8)  & & & & 37 & (21,16) & & & & 53 & (32,21) \\ 
\multicolumn{1}{ c }{6} & (4,2) & & &  &  22 & (14,8)  & & & & 38 & (22,16) & & & & 54 & (32,22) \\ 
\multicolumn{1}{ c }{7} & (4,3) & & &  & 23  & (15,8)  & & & & 39 & (23,16) & & & & 55 & (32,23) \\ 
\multicolumn{1}{ c }{8} & (4,4) & & &  &  24 & (16,8)  & & & & 40 & (24,16) & & & & 56 & (32,24) \\ 
\multicolumn{1}{ c }{9} & (5,4) & & &  &  25 & (16,9)  & & & & 41 & (25,16) & & & & 57 & (32,25) \\ 
\multicolumn{1}{ c }{10} & (6,4) & & &  &  26 & (16,10)  & & & & 42 & (26,16) & & & & 58 & (32,26) \\ 
\multicolumn{1}{ c }{11} & (7,4) & & &  & 27  & (16,11)  & & & & 43 & (27,16) & & & & 59 & (32,27) \\ 
\multicolumn{1}{ c }{12} & (8,4) & & &  & 28  & (16,12)  & & & & 44 & (28,16) & & & & 60 & (32,28) \\ 
\multicolumn{1}{ c }{13} & (8,5) & & &  &  29 & (16,13)  & & & & 45 & (29,16) & & & & 61 & (32,29) \\ 
\multicolumn{1}{ c }{14} & (8,6) & & &  & 30 & (16,14)  & & & & 46 & (30,16) & & & & 62 & (32,30) \\ 
\multicolumn{1}{ c }{15} & (8,7) & & &  &  31 & (16,15)  & & & & 47 & (31,16) & & & & 63 & (32,31) \\ 
\multicolumn{1}{ c }{16} & (8,8) & & &  &  32 & (16,16)  & & & & 48 & (32,16) & & & & 64 & (32,32) \\
\multicolumn{1}{ c }{17} & (9,8) & & &  &  33 & (17,16)  & & & & 49 & (32,17) & & & & 65 & (33,32) \\\cmidrule{1-17}
\end{tabular}
\label{table:leftright}
\end{center}
Note. --- The tree with the maximum number of rankings splits into (left, right) subtrees with $(\ell,r)$ leaves. The $n$-taxon species tree with the maximum number of rankings $T_n$ has $2^{1+\floor*{\log_2[(n-1)/3]}}$ taxa descended from one side of the root and $n-2^{1+\floor*{\log_2[(n-1)/3]}}$ from the other side.
\end{table}

\par
\begin{table}[htbp]
\footnotesize
\begin{center}
\caption{The number of balanced internal vertices $\sigma(T_n)$ in $n$-taxon species trees with the maximum number of rankings for a labeled topology.}
\label{table:symtable}
\begin{tabular}{r r r r  r r r}
\cmidrule[0.5pt]{1-7}
\multicolumn{3}{c}{$n$ even} & \multicolumn{1}{c}{} & \multicolumn{3}{c}{$n$ odd} \\\cmidrule{1-7}
\multicolumn{1}{c}{$n$}&\multicolumn{1}{c}{$\sigma(T_n)$}&\multicolumn{1}{c}{$n-1-\sigma(T_n)$}&\multicolumn{1}{c}{}&\multicolumn{1}{c}{$n$}&\multicolumn{1}{c}{$\sigma(T_n)$}&\multicolumn{1}{c}{$n-1-\sigma(T_n)$}\\\midrule
\multicolumn{1}{ c }{2} & 1 & 0 & & 3 & 1  & 1 \\ 
\multicolumn{1}{ c }{4} & 3 & 0 & & 5 & 2  & 2 \\ 
\multicolumn{1}{ c }{6} & 4 & 1 & & 7 & 4  & 2 \\\cmidrule{1-7}
\multicolumn{1}{ c }{8} & 7 & 0 & & 9 & 5  & 3 \\ 
\multicolumn{1}{ c }{10} & 7 & 2 & & 11 & 7 & 3 \\ 
\multicolumn{1}{ c }{12} & 10 & 1 & & 13 & 9 & 3 \\ 
\multicolumn{1}{ c }{14} & 11 & 2 & & 15 & 11 & 3 \\\cmidrule{1-7}
\multicolumn{1}{ c }{16} & 15 & 0 & & 17 & 12 & 4 \\ 
\multicolumn{1}{ c }{18} & 14 & 3 & & 19 & 14 & 4 \\ 
\multicolumn{1}{ c }{20} & 17 & 2 & & 21 & 16 & 4 \\ 
\multicolumn{1}{ c }{22} & 18 & 3 & & 23 & 18 & 4 \\ 
\multicolumn{1}{ c }{24} & 22 & 1 & & 25 & 20 & 4 \\ 
\multicolumn{1}{ c }{26} & 22 & 3 & & 27 & 22 & 4 \\ 
\multicolumn{1}{ c }{28} & 25 & 2 & & 29 & 24 & 4 \\ 
\multicolumn{1}{ c }{30} & 26 & 3 & & 31 & 26 & 4 \\\cmidrule{1-7}
\multicolumn{1}{ c }{32} & 31 & 0 & & 33 & 27 & 5 \\ 
\multicolumn{1}{ c }{34} & 29 & 4 & & 35 & 29 & 5 \\ 
\multicolumn{1}{ c }{36} & 32 & 3 & & 37 & 31 & 5 \\ 
\multicolumn{1}{ c }{38} & 33 & 4 & & 39 & 33 & 5 \\ 
\multicolumn{1}{ c }{40} & 37 & 2 & & 41 & 35 & 5 \\ 
\multicolumn{1}{ c }{42} & 37 & 4 & & 43 & 37 & 5 \\ 
\multicolumn{1}{ c }{44} & 40 & 3 & & 45 & 39 & 5 \\ 
\multicolumn{1}{ c }{46} & 41 & 4 & & 47 & 41 & 5 \\ 
\multicolumn{1}{ c }{48} & 46 & 1 & & 49 & 43 & 5 \\ 
\multicolumn{1}{ c }{50} & 45 & 4 & & 51 & 45 & 5 \\ 
\multicolumn{1}{ c }{52} & 48 & 3 & & 53 & 47 & 5 \\ 
\multicolumn{1}{ c }{54} & 49 & 4 & & 55 & 49 & 5 \\ 
\multicolumn{1}{ c }{56} & 53 & 2 & & 57 & 51 & 5 \\ 
\multicolumn{1}{ c }{58} & 53 & 4 & & 59 & 53 & 5 \\ 
\multicolumn{1}{ c }{60} & 56 & 3 & & 61 & 55 & 5 \\ 
\multicolumn{1}{ c }{62} & 57 & 4 & & 63 & 57 & 5 \\\cmidrule{1-7}
\multicolumn{1}{ c }{64} & 63 & 0 & & 65 & 58 & 6 \\ 
\end{tabular}
\end{center}
Note. --- For even $n$, $\sigma(T_n) \geq n-k-1$ (Lemma 3). For completely balanced and symmetric $n=2^{k+2}$-taxon trees, $\sigma(T_n) = n-1$. For $n=3 \cdot 2^{\floor*{\log_2(n)-1}}$-taxon trees, $\sigma(T_n) = n - 2$. For odd $n$, the number of balanced internal vertices is $\sigma(T_n) = n-1-\floor*{\log_2 n}$ (Lemma 2).
\end{table}

\end{document}